\documentclass[fleqn,usenatbib,useAMS]{mnras}

\usepackage{graphicx}	
\usepackage{amsmath}	
\usepackage{amssymb}	
\usepackage{multicol}        
\usepackage{bm}		
\usepackage{pdflscape}	
\usepackage{xcolor}

\newcommand{\HI}{\ion{H}{i}}
\newcommand{\HII}{\ion{H}{ii}}
\newcommand{\Hmol}{H$_2$}

\newcommand{\mhi}{M_{\rm HI}}
\newcommand{\kpc}{\,{\rm kpc}}
\newcommand{\mpc}{\,{\rm Mpc}}

\newcommand{\hmpc}{\,h^{-1}{\rm Mpc}}

\newcommand{\Msun}{\,\rm {\rm M}_\odot}

\newcommand{\kms}{\;{\rm km}\,{\rm s}^{-1}}

\newcommand{\msolar}{\;{\rm M}_{\odot}}

\newcommand{\mufasa}{{\sc Mufasa}}
\newcommand{\simba}{{\sc Simba}}

\newcommand{\fneut}{f_{\rm HI}}
\newcommand{\fmol}{f_{\rm H_2}}
\newcommand{\fedd}{f_{\rm Edd}}

\newcommand{\TNG}{IllustrisTNG}
\newcommand{\alphaCO}{\alpha_{\rm CO}}
\newcommand{\aCOunits}{{\rm M}_\odot {\rm pc}^{-2} {\rm K}^{-1} {\rm km}^{-1} {\rm s}}

\newcommand{\romeel}[1]{\color{black}{#1}\color{black}}
\newcommand{\romeelnew}[1]{\color{black}{#1}\color{black}}
\newcommand{\adam}[1]{\color{black}{#1}\color{black}}

\usepackage[T1]{fontenc}
\usepackage{ae,aecompl}

\usepackage{newtxtext,newtxmath}

\title[Cold Gas in Hydro Simulations]{Galaxy Cold Gas Contents in Modern Cosmological Hydrodynamic Simulations}

\author[Dav\'e et al.]{%
Romeel Dav\'e,$^{1,2,3}$ 
Robert A. Crain,$^{4}$
Adam R.~H.~Stevens,$^{5,6}$
Desika Narayanan,$^{7,8,9}$
\newauthor
Amelie Saintonge,$^{10}$
Barbara Catinella,$^{5,6}$
Luca Cortese$^{5,6}$
\\
\\$^1$ Institute for Astronomy, Royal Observatory, Univ. of Edinburgh, Edinburgh EH9 3HJ, UK
\\$^2$ University of the Western Cape, Bellville, Cape Town 7535, South Africa
\\$^3$ South African Astronomical Observatories, Observatory, Cape Town 7925, South Africa
\\$^4$ Astrophysics Research Institute, Liverpool John Moores University, 146 Brownlow Hill, Liverpool, L3 5RF
\\$^5$ International Centre for Radio Astronomy Research, The University of Western Australia, Crawley, WA 6009, Australia
\\$^6$ ARC Centre of Excellence for All Sky Astrophysics in 3 Dimensions (ASTRO 3D), Australia
\\$^7$ Department of Astronomy, University of Florida, 211 Bryant Space Sciences Center, Gainesville, FL, USA
\\$^8$ University of Florida Informatics Institute, 432 Newell Drive, CISE Bldg E251, Gainesville, FL, USA
\\$^9$ Cosmic Dawn Center at the Niels Bohr Institute, University of Copenhagen and DTU-Space, Technical University of Denmark
\\$^{10}$ Department of Physics and Astronomy, University College London, Gower Street, London WC1E 6BT, UK
}

\date{Accepted XXX. Received YYY; in original form ZZZ}

\pubyear{2019}

\begin{document}
\label{firstpage}
\pagerange{\pageref{firstpage}--\pageref{lastpage}}
\maketitle

\begin{abstract}
We present a comparison of galaxy atomic and molecular gas properties in three recent cosmological hydrodynamic simulations, \simba, EAGLE, and \TNG, versus observations from $z\sim 0-2$. These simulations \romeel{all rely on similar sub-resolution prescriptions to model cold interstellar gas which they cannot represent directly, and qualitatively}~reproduce the observed $z\approx 0$ \HI\ and H$_2$ mass functions (HIMF, H2MF), CO(1-0) luminosity functions (COLF), and gas scaling relations versus stellar mass, specific star formation rate, and stellar surface density $\mu_*$, \romeel{with some quantitative differences}.  \romeel{To compare to the COLF, we apply an H$_2$-to-CO conversion factor to the simulated galaxies based on their average molecular surface density and metallicity,}~yielding substantial variations in $\alphaCO$ and significant differences between models.  Using this, predicted $z=0$ COLFs agree better with data than predicted H2MFs. Out to $z\sim 2$, EAGLE's and \simba's HIMF and COLF strongly increase, while \TNG's HIMF declines and COLF evolves slowly.  EAGLE and \simba\ reproduce high $L_{\rm CO1-0}$ galaxies at $z\sim 1-2$ as observed, owing partly to a median $\alphaCO(z=2)\sim 1$ versus $\alphaCO(z=0)\sim 3$. Examining \HI, H$_2$, and CO scaling relations, their trends with $M_*$ are broadly reproduced in all models, but EAGLE yields too little \HI\ in green valley galaxies, \TNG\ and \simba\ overproduce cold gas in massive galaxies, and \simba\ overproduces molecular gas in small systems.  Using \simba\ variants that exclude individual AGN feedback modules, we find that \simba's AGN jet feedback is primarily responsible by lowering cold gas contents from $z\sim 1\to0$ by suppressing cold gas in $M_*\ga 10^{10}{\rm M}_\odot$ galaxies, while X-ray feedback suppresses the formation of high-$\mu_*$ systems. 
\end{abstract}




\romeel{}

\section{Introduction}

Galaxies are made up of stars, gas, dust, black holes, and dark matter.  Of those, stars represent the most straightforwardly visible component, and so have received the most observational and theoretical attention.  But in the modern baryon cycling paradigm of galaxy evolution, it is the exchange of gas between the interstellar medium (ISM) of galaxies and their surrounding circum-galactic medium (CGM) via inflows, outflows, and recycling that primarily govern how galaxies form and evolve~\citep{Tumlinson:2017}.  As such it is becoming clear that understanding the gas within and around galaxies is crucial for a full picture of galaxy formation and evolution.

Comprehensive observations of gas in galaxies are challenging because gas is typically diffuse, multi-phase, and its emission is often best traced in less accessible portions of the electromagnetic spectrum. For instance, molecular gas (primarily H$_2$) is canonically traced via heavy element molecular emission lines in the millimetre regime; atomic gas is most straightforwardly traced as radio 21-cm emission; ionised gas appears as ultraviolet and optical emission lines; and gas at high temperatures is mostly evident via X-ray emission.  Assembling observations of these various gas phases into a coherent scenario for the role of gas in galaxy evolution is an important goal for current models of galaxy evolution.

Within the interstellar medium (ISM) of star-forming galaxies, the dominant gas phases are cold ($T\la 100$~K) and warm ($T\sim 10^3-10^4$~K), best traced by molecular and atomic hydrogen, respectively.  Recent advances in observations of molecular and atomic gas have opened up new windows on understanding the role of ISM gas in galaxy evolution.  This gas not only provides the reservoir for new star formation, but also contains strong signatures of feedback processes from both star formation (SF) and active galactic nuclei (AGN).  Hence it is important to situate ISM gas in and around galaxies within the context of hierarchical galaxy formation.

Cosmological gas dynamical simulations provide a comprehensive approach towards elucidating the connection between ISM gas, star formation, and feedback.  Modern simulations now include sophisticated models for star formation and feedback processes, and generally do a good job of reproducing the observed evolution of the stellar component, including suppressing low-mass galaxy growth via stellar feedback and producing massive quenched galaxies via AGN feedback~\citep{Somerville:2015}.  Simulations such as EAGLE~\citep{Schaye:2015}, \TNG~\citep{Pillepich:2018}, and \simba~\citep{Dave:2019} all produce stellar mass functions in reasonable agreement with observations over cosmic time, showing a shallow faint-end slope and a truncation at high masses coincident with the onset of a quiescent galaxy population.  Despite this concordance, the detailed physical models for sub-grid processes such as star formation and feedback implemented in each model are markedly different. Hence discrimination between such simulations requires comparing to data beyond stellar masses and stellar growth rates.

Emerging observations of cold ISM gas provide a new regime for testing galaxy formation models. Early simulations demonstrated a clear connection between SF-driven feedback and the cold gas contents of galaxies.  For instance, \citet{Dave:2011b} showed, perhaps counter-intuitively, that increasing the strength of galactic outflows results in an increased gas fraction at a a fixed stellar mass in galaxies; while the cold gas mass at a given halo mass is lower, the stellar mass is reduced even further~\citep{Crain:2017}.  The \mufasa\ simulation used an improved model for SF feedback and obtained good agreement with observations available at the time~\citep{Dave:2017a}.  \romeel{EAGLE~\citep{Schaye:2015,Crain:2015} has been successful in reproducing a wide range of galaxy properties, and has been used to investigate the origin of \HI\ in galaxies~\citep{Crain:2017}}.
\TNG\ has been shown to broadly capture many observed cold-gas statistics of galaxies, including trends with galaxy environment, but some curious features in the gas content of massive galaxies that are likely related to AGN feedback remain in tension with observations~\citep{Diemer:2018,Stevens:2019}.  Despite the increased uncertainty with modeling the observational characteristics of gas in simulated galaxies, and the limited resolution that precludes direct modeling of many detailed ISM processes in cosmological volumes, these results highlight how cold gas observations could potentially provide a valuable testbed for modern galaxy formation models.

Observations of cold gas components within galaxies have also matured in recent years.  For atomic hydrogen, large blind \HI\ surveys such as the \HI\ Parkes All-Sky Survey~\citep[HIPASS;][]{Barnes:2001,Meyer:2004,Wong:2006} and the Arecibo Legacy Fast ALFA survey~\citep[ALFALFA;][]{Giovanelli:2005,Haynes:2018} characterised the properties of \HI-selected galaxies over a wide area, but since these galaxies were selected by their \HI\ mass, this meant that these surveys tended to preferentially detect \HI-rich systems~\citep{Catinella:2012}.  In order to connect to models, it is more optimal to have a survey that selects on a quantity that is more robustly predicted in models.  Ideally, this is stellar mass, since it is the observation that models are most commonly tuned to reproduce, enabling a equal-footing comparison between simulation predictions.

The GALEX Arecibo SDSS Survey~\citep[GASS;][]{Catinella:2010} was pioneering in that it measured \HI\ contents for a stellar mass selected sample from the Sloan Digital Sky Survey with $M_*\ga 10^{10}{\rm M}_\odot$, using Arecibo.  GASS was able to statistically quantify or place limits on \HI-poor galaxies, which become increasingly commonplace towards higher masses.  To expand the dynamic range, the GASS-Low survey was done to extend the completeness down to $M_*\ga 10^{9}{\rm M}_\odot$. The aggregate survey, known as extended GASS (xGASS), thus provides \HI\ contents and upper limits for a representative sample of nearly 1200 galaxies with $10^9<M_*< 10^{11.5}{\rm M}_\odot$~\citep{Catinella:2018}.

Molecular gas measurements have likewise made significant progress in recent years, typically via CO surveys. As a complement to xGASS, the xCOLD GASS survey~\citep{Saintonge:2011,Saintonge:2017} provided CO(1-0) and CO(2-1) measurements for over 500 of the xGASS-observed galaxies.  As with xGASS, the careful selection from SDSS enabled reconstruction of a volume-limited sample. xGASS and xCOLD GASS thus provide a benchmark constraint for modern cosmological galaxy formation models, with a well-specified selection function that allows cleaner model--data comparisons, and comprehensive ancillary data that enables tests of the relationship between the cold gas and other components of galaxies.

Moving to higher redshifts, \HI\ surveys are currently quite limited~\citep{Catinella:2015,Fernandez:2016}, though for example the recently begun LADUMA survey on MeerKAT aims to measure \HI\ directly out to $z\ga 1$~\citep{Blyth:2016}.  CO surveys at higher redshifts meanwhile have progressed substantially.  The pioneering Plateau de Bure HIgh-z Blue Sequence Survey (PHIBBS) survey was able to study the molecular content of a well-studied sample including internal kinematics~\citep{Tacconi:2013} to $z\sim 2$, but this was not designed to sample a representative volume.   The recent CO Luminosity Density at High-z (COLDz) survey using the Jansky Very Large Array~\citep{Pavesi:2018} and the Atacama Large Millimetre Array (ALMA) SPECtroscopic Survey (ASPECS) in the Hubble Ultra Deep Field~\citep{Aravena:2019} have provided a more statistical characterisation of the molecular gas contents of galaxies out to $z\sim 2$, albeit with limited samples.  Given that such high redshift cold gas observations are set to improve dramatically in the next few years, the time is ripe to provide a snapshot view of how modern galaxy formation simulations that are successful in reproducing stellar properties fare against available cold gas observations.

This paper compares the predictions from three recent cosmological hydrodynamic simulations, namely \simba, EAGLE, and IllustrisTNG, to ALFALFA, xGASS, and xCOLD GASS observations at low redshifts, and ASPECS and COLDz data at high redshifts.  The primary purpose is to assess the range of predictions among state-of-the-art hydrodynamic models in galaxy cold gas contents, and to provide preliminary comparisons to observations.  A proper comparison would involve mimicking details observational selection effects for each survey, which we leave for future work; here we take extant simulation predictions for EAGLE and \TNG, add \simba\ predictions, and compare to data at face value.  We examine \HI\ and H$_2$ contents that are more directly predicted in these simulations, along with the CO(1-0) luminosity determined via a conversion factor following the recipe based on merger simulations and CO radiative transfer by \citet{narayanan12a}.  We find substantial variations among current models in their cold gas predictions, with all models qualitatively reproducing the broad trends but no model quantitatively reproducing all the observations. 

This paper is organised as follows.  In \S\ref{sec:sims} we briefly recap the simulations and key observations used in this work.  In \S\ref{sec:results} we first compare stellar properties to show that they are similar among our simulations, and then present our comparisons from $z=0-2$ for the \HI\ and H$_2$ mass functions, the H$_2$-to-CO conversion factor, the CO(1-0) luminosity functions, and the gas content scaling relations at $z=0$, all compared to a range of relevant observations focused on xGASS and xCOLD~GASS at $z\sim 0$ along with other recent surveys from $z\sim 0-2$.  We further examine different variants of AGN feedback models within \simba\ to better understand the physics driving the evolution of cold gas mass functions.  Finally, in \S\ref{sec:summary} we summarise our results.

\section{Simulations \& Observations}\label{sec:sims}

In this work we employ the \simba\, EAGLE, and \TNG\ simulations, and will compare these to the xGASS and xCOLD~GASS data sets.  Here we briefly review these simulations and observations.

\romeel{Beyond briefly describing each simulation's input physics, we will focus on each one's procedure for partitioning gas into ionised, atomic, and molecular phases.  As cosmological simulations, each of these models has a spatial resolution of $\sim 1$kpc, which is insufficient to directly model physical processes giving rise to the cold ISM phase.  Even the onset to self-shielding is not done self-consistently, since this would require radiative transfer, which has a prohibitive computational cost.  Instead, each model employs a set of established but approximate sub-grid prescriptions in order to determine the atomic and molecular fractions of dense gas.  The prescription for self-shielding to form neutral gas is essentially the same among these models, while that for forming molecular gas is not identical but still broadly similar.}

\subsection{\simba}\label{sec:simba}

\simba~\citep{Dave:2019},is the successor to the \mufasa\ simulation~\citep{Dave:2016}, which was run using a modified version of the gravity plus hydrodynamics solver \textsc{Gizmo} \citep{Hopkins:2015} in its meshless finite mass (MFM) mode. The simulation evolves a representative 100$\hmpc$ comoving volume from $z=249\to 0$ with $1024^3$ gas elements and $1024^3$ dark matter particles.  The mass resolution is \(9.6\times 10^7 M_{\odot}\) for dark matter particles and \(1.82\times 10^7 M_{\odot}\) for gas elements, and the minimum adaptive gravitational softening length is \(\epsilon_{min} = 0.5h^{-1}\)ckpc.  Initial conditions are generated using \textsc{Music} \citep{Hahn:2011} assuming the following cosmology~\citep{Planck:2016}: \(\Omega_M = 0.3\), \(\Omega_{\Lambda} = 0.7\), \(\Omega_{b} = 0.048\), \(H_0 = 68\) km s\(^{-1}\) Mpch\(^{-1}\), \(\sigma_8 = 0.82\), \(n_s = 0.97\). 

Star formation is modelled using an H$_2$-based \citet{Schmidt:1959} relation, where the H$_2$ fraction is computed using the sub-grid prescription of \citet{Krumholz:2011} based on the local metallicity and gas column density, modified to account for variations in resolution \citep{Dave:2016}. \romeel{This H$_2$ fraction will be directly used in the H$_2$ results for this paper, so bears further description.  For each gas element, the H$_2$ fraction is computed as
\begin{equation}
f_{\rm H2} = 1 - 0.75{s\over 1+0.25s}
\end{equation}
where
\begin{equation}
s = {\ln (1+0.6\chi+0.01\chi^2)\over 0.0396 Z (\Sigma/M_\odot {\rm pc}^{-2})},
\end{equation}
where $Z$ is the metallicity in solar units, $\chi$ is a function
of metallicity \citep[see][]{Krumholz:2011}, and $\Sigma=\rho^2/|\nabla\rho|$ is the column density calculated using the Sobolev approximation, increased by $\times 30^{2/3}$ to account for sub-resolution clumping~\citep[see][for full discussion]{Dave:2016}. We impose a minimum metallicity of $10^{-3}Z_\odot$ solely for the
purposes of this sub-grid model. 
}

The star formation rate (SFR) is calculated from the density $\rho$ and the dynamical time \(t_{\textrm{dyn}}\) via \(\textrm{SFR}=\epsilon_* f_{\rm H_2}\rho/t_{\textrm{dyn}} \), where \(\epsilon_* = 0.02\) \citep{Kennicutt:1998}.  \romeel{Star formation is only allowed to occur in gas with $n_H>0.13$~H~atoms~cm$^{-3}$, although the limiting factor is $\fmol>0$ for all but super-solar metallicities.  \simba\ artificially pressurises  gas above this density by imposing $T=10^4 (n_H/0.13)^{4/3}$~K~\citep{Schaye:2008}, in order to prevent numerical fragmentation owing to the Jeans mass being unresolved.}

The \HI\ fraction of gas elements is computed self-consistently within the code, accounting for self-shielding on the fly based on the prescription in \citet{Rahmati:2013}, where the metagalactic ionizing flux strength is attenuated depending on the gas density assuming a spatially uniform ionising background from \citet{Haardt:2012}.  This gives the total neutral gas, and subtracting off the H$_2$ yields the \HI.  Hence in \simba, the \HI\ and H$_2$ fractions for gas are computed self-consistently, on the fly during the simulation run.

\begin{figure}
	\includegraphics[width=0.5\textwidth]{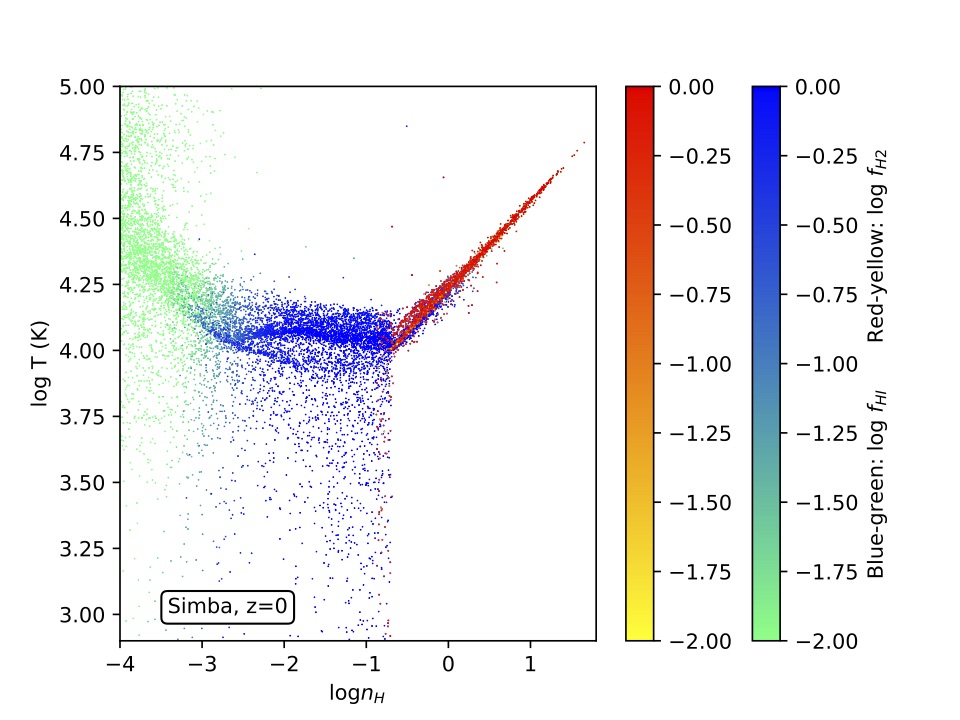}
	\vskip-0.1in
    \caption{Phase diagram of hydrogen number density $n_H$ vs. temperature $T$ for dense, cool gas in \simba\ at $z=0$. For clarity, a random sample of 0.05\% of gas elements are plotted.  Points are colour-coded by \HI\ fraction in blue/green, and H$_2$ fraction in red/yellow for the gas elements where $\fmol>0$. Abrupt transitions are seen from nearly fully ionised to nearly fully neutral at $n_H\sim 10^{-3}$ cm$^{-3}$, while the transition from atomic to molecular is similarly abrupt but depends on metallicity.}
    \label{fig:rhot_dense}
\end{figure}

\romeel{Figure~\ref{fig:rhot_dense} illustrates the resulting cold gas fractions in phase space in \simba\ at $z=0$.  A random subsample of 0.05\% of all particles are plotted in $(n_H,T)$ space, with green$\to$blue colours showing the \HI\ fractions, and yellow$\to$red colors indicating H$_2$ fractions in gas where $\fmol>0$.  This plot focuses on the cool, dense phase of cosmic gas, since other cosmic phases have tiny neutral fractions; see \citet{Christiansen:2019} for a complete phase diagram from \simba.  For EAGLE, the analogous diagram is shown in Figure~1 of \citet{Crain:2017}.

There is an abrupt transition at $n_H\approx 10^{-3}$~cm$^{-3}$ above which self-shielding kicks in, and the gas goes from highly ionised to mostly neutral.  Metal cooling can cool highly enriched gas to fairly low temperatures, but above $n_H>0.13$~cm$^{-3}$, the ISM pressurisation becomes evident as the gas is forced to have a minimum Jeans mass-resolving temperature of $T\propto\rho^{1/3}$~\citep{Dave:2016}. Ultimately, it is this ISM pressurisation forced by the poor resolution~\citep{Schaye:2008} that prevents direct modeling of the cold molecular ISM phase in \simba, as well as in the other cosmological simulations.  In this pressurized region, the gas fairly abruptly transition to molecular, but the density at which this occurs is metal-dependent, hence there is still substantial low-metallicity atomic gas above the threshold (blue points, many hidden underneath the red points).
This illustrates how self-shielding and the \citet{Krumholz:2011} prescription work together to transform ionised IGM gas into atomic and molecular phases. }

Radiative cooling and photoionisation heating are implemented using the \textsc{Grackle}-3.1 library \citep{Smith:2017}. The chemical enrichment model tracks 9 metals during the simulation, tracking enrichment from Type II supernovae (SNe), Type Ia SNe and asymptotic giant branch (AGB) stars, including locking some of the metals into dust. A \citep{Chabrier:2003} stellar initial mass function is assumed in order to compute stellar evolution.  \simba\ includes star formation-driven galactic winds as kinetic two-phase, metal-enriched winds with 30\% of the wind particles ejected hot and with a mass loading factor that scales with stellar mass, based on the FIRE \citep{Hopkins:2014} zoom simulation scalings from \citet{Angles:2017b}.  Importantly, hydrodynamics are turned off in the winds (``decoupled'') until they are well outside the interstellar medium~\citep{Springel:2003}, hence they explicitly avoid depositing energy in the ISM on their way out.  This is done because in any current hydrodynamic solver, a single fluid element moving at high velocities through an ambient medium is more accurately represented by turning off hydrodynamics rather than allowing the solver to calculate the interactions.

\simba's main improvement on \mufasa\ is the addition of black hole growth via torque-limited accretion, and AGN feedback via stable bipolar kinetic outflows.  For (\(T <10^5\) K) gas, black hole accretion follows the torque limited accretion model of \citet{Angles:2017} which is based on the analytic model of \citet{Hopkins:2011}, while for hot gas accretion \citep{Bondi:1952} accretion is employed. 
AGN feedback in \simba\ is designed to mimic the observed dichotomy in black hole growth modes seen in real AGN \citep[e.g.][]{Heckman:2014}: a `radiative' mode at high Eddington ratios ($\fedd$) characterised by mass-loaded radiatively-driven winds ejected at $\sim 10^3\kms$, and a `jet' mode at low $\fedd\la$ few percent at $\sim 10^4\kms$. The mass loading is set such that the outflow momentum is $20L/c$, where $L=0.1\dot{M}c^2$ is the radiative luminosity for a black hole accretion rate of $\dot{M}$.  Additionally, we include X-ray heating by black holes based on the model of \citet{Choi:2012}.  This yields a quenched galaxy population~\citep{Rodriguez:2019} and galaxy--black hole co-evolution~\citep{Thomas:2019} in good agreement with observations. 

In addition to the full \simba\ feedback model, we will consider $50\hmpc$, $2\times 512^3$ variants where we turn off the X-ray feedback (``No-X''), turn off both X-ray and jet feedback (``No-jet''), and turn off all AGN feedback (``No-AGN'').  These runs have the same resolution as the $100\hmpc$ run but with $1/8$ the volume.  All other input physics, as well as the initial conditions, are identical. These models will be described further in \S\ref{sec:simba_agn}. Finally, to assess resolution convergence, we will employ a $25\hmpc$, $2\times 512^3$ \simba\ run, with identical input physics to the full \simba\ run. This has $8\times$ better mass resolution than the other \simba\ runs. Feedback parameters have not been re-tuned at this higher resolution.

Galaxies are identified using a 6-D friends-of-friends (FOF) galaxy finder, using a spatial linking length of 0.0056 times the mean inter-particle spacing (equivalent to twice the minimum softening kernel), and a velocity linking length set to the local velocity dispersion.  This is applied to all stars and ISM gas ($n_H>0.13$~cm$^{-2}$).
Halos are identified using a 3-D FOF with a linking length parameter of 0.2.  The \HI\ and H$_2$ fractions for individual gas elements are taken directly from the simulation, without any post-processing.  To compute galaxies' \HI\ and H$_2$ contents, \romeel{we assign each gas particle in a halo to the galaxy that has the highest value of $M_{\rm baryon}/R^2$, where $M_{\rm baryon}$ is the total baryonic mass of the galaxy, and $R$ is the distance from the particle to the galaxy's centre of mass.  This enables cold gas, particularly \HI, to be assigned to a galaxy even if it is not identified as within the galaxy's ISM, since the \HI\ can be significant even for gas with $n_H<0.13$~cm$^{-2}$.  For H$_2$, the results are insensitive to whether we consider this low-density gas, since all the molecular gas is located within the ISM (Fig.~\ref{fig:rhot_dense}).
}

For consistency among the models, we will restrict our simulated galaxy samples to $M_*>10^9\; {\rm M}_\odot$, which represents the approximate mass limit for the primary observational sample to which we will compare, even though all our simulations are able to resolve to lower stellar masses.

\subsection{EAGLE}

The EAGLE simulations \citep[][with public data release described by \citealt{McAlpine:2016}]{Schaye:2015, Crain:2015} were evolved with a substantially-modified version of the $N$-body Tree-Particle-Mesh (TreePM) smoothed particle hydrodynamics (SPH) solver \textsc{gadget3}, \citep[last described by][]{Springel:2005}. The modifications include significant updates to the hydrodynamics solver, the time-stepping criteria, and the implemented sub-grid physics modules. The largest-volume EAGLE simulation \citep[Ref-L100N1504 in the nomenclature of][]{Schaye:2015} evolves a region $67.77\hmpc$ ($100\mpc$) to the present day, realised with $1504^3$ dark matter particles and an (initially) equal number of baryonic particles, yielding particle masses of $1.81 \times 10^6 \Msun$ and $9.70 \times 10^6 \Msun$ for baryons and dark matter, respectively. The Plummer-equivalent gravitational softening length is $\epsilon_{\rm com} = 2.66\,{\rm ckpc}$, limited to a maximum proper length of $\epsilon_{\rm prop} = 0.7\,{\rm pkpc}$. We will also show results from the EAGLE-Recal simulation \citep[Recal-L25N752 in][]{Schaye:2015}, which evolved a $16.94\hmpc$ volume with $8\times$ higher mass resolution, and $2\times$ higher spatial resolution. Initial conditions were generated with the software described by \citet{Jenkins:2010}, as detailed in Appendix B of \citet{Schaye:2015}, assuming the \citet{Planck:2013} cosmogony: \(\Omega_{\rm m} = 0.307\), \(\Omega_{\Lambda} = 0.693\), \(\Omega_{\rm b} = 0.04825\), \(H_0 = 67.77\) km s\(^{-1}\) Mpc\(^{-1}\), \(\sigma_8 = 0.8288\), \(n_s = 0.9611\). 

Interstellar gas is treated as a single-phase star-forming fluid with a polytropic pressure floor \citep{Schaye:2008}, subject to a a metallicity-dependent density threshold for star formation \citep{Schaye:2004}, which reproduces (by construction) the observed Kennicutt-Schmidt relation \citep{Kennicutt:1998} in gas that satisfies local vertical hydrostatic equilibrium. Radiative heating and cooling are implemented element-by-element for 11 species \citep{Wiersma:2009a} in the presence of a time-varying UV/X-ray background radiation field \citep{HaardtMadau:2001} and the cosmic microwave background (CMB). The evolution of the same species due to stellar evolution and mass loss are tracked during the simulation according to the implementation of \citet{Wiersma:2009b}. The seeding of BHs and their growth via gas accretion and BH-BH mergers are treated with an updated version of the method introduced by \citet{SpringelDiMatHern:2005}, accounting for the dynamics of gas close to the BH \citep{Rosas:2015}. Feedback associated with the formation of stars \citep{DallaVecchia:2012} and the growth of BHs \citep{Booth:2009} are both implemented via stochastic, isotropic heating of gas particles ($\Delta T_{\rm SF} = 10^{7.5}$ K, $\Delta T_{\rm AGN} = 10^{8.5}$ K), designed to prevent immediate, numerical radiative losses. Heated particles are not decoupled from the hydrodynamics scheme. The simulations assume the \citet{Chabrier:2003} IMF.

Halos are identified by applying the friends-of-friends (FoF) algorithm to the dark matter particle distribution, with a linking length of 0.2 times the mean inter-particle separation. Gas, stars and BHs are associated with the FoF group, if any, of their nearest dark matter particle. Galaxies are equated to bound substructures within halos, identified by the application of the {\sc Subfind} algorithm \citep{Springel:2001,Dolag:2009} to the particles (of all types) comprising FoF haloes. Following \citet{Schaye:2015}, we compute the properties of galaxies by aggregating the properties of the relevant particles residing within $30\kpc$ of their most-bound particle. 

The EAGLE model does not partition hydrogen into its ionised (\HII), atomic (\HI) and molecular (\Hmol) forms `on-the-fly', so we estimate the fraction of each on a particle-by-particle basis in post-processing, with the two-step approximation used by \citet[][elements of which were also used by \citealt{Lagos:2015,Lagos:2016,Bahe:2016,Marasco:2016}]{Crain:2017}. Hydrogen is first partitioned into atomic (\HI\ + \Hmol) and ionized (\HII) components, using the fitting function of \citet{Rahmati:2013}, which considers both collisional ionization \citep[using temperature-dependent rates collated by][]{Cen:1992} and photoionization by metagalactic UV radiation, calibrated using TRAPHIC radiative transfer simulations \citep{Pawlik:2008}. Radiation due to sources within or local to galaxies, although likely significant \citep[see e.g.][]{MiraldaEscude:2005} is not considered explicitly, but is accounted for implicitly via the use of an empirical or theoretical scheme to partition the atomic hydrogen of star-forming gas particles into its atomic and molecular components. 
\citet{Crain:2017} present results using two such schemes, to illustrate the associated systematic uncertainty. The first is the theoretically-motivated prescription of \citet{GK:2011}, whilst the second is motivated by the observed scaling of the molecular to atomic hydrogen surface density ratio ($R_{\rm mol} \equiv \Sigma_{{\rm H}_2}/\Sigma_{{\rm HI}}$) with the mid-plane pressure of galaxy discs \citep{BR:2006}. In the latter case, gas particles are assigned a molecular hydrogen fraction that scales as a function of their pressure \citep[this approach has been widely used elsewhere, see e.g.][]{Popping:2009,Duffy:2012,Dave:2013,Rahmati:2013}. 

\romeel{For the EAGLE results in this paper, we will employ the results using the \citet{GK:2011} approach. This is primarily because it is significantly closer to the approaches used in \simba\ and \TNG.}~ \citet{Crain:2017} show that this scheme generally yields higher molecular fractions than the pressure-based approach, and that the partitioning of neutral hydrogen into its atomic and molecular components is a more severe uncertainty on the \HI\ masses of galaxies than, for example, the freedom afforded by present constraints on the amplitude of the present-day UV background. However, the primary systematic influence on atomic hydrogen masses was found to be the resolution of the simulation, with galaxies of fixed mass exhibiting significantly greater ISM masses when simulated at 8x (2x) greater mass (spatial) resolution, \romeel{as will be reiterated in the results of this paper.}


\subsection{\TNG}

IllustrisTNG comprises a suite of cosmological magnetohydrodynamic simulations at various resolutions and volumes, all run with the moving-mesh {\sc arepo} code~\citep{Springel:2010_arepo}, assuming a $\Lambda$CDM cosmology with parameters based on ~\citet{Planck:2016}, using a common galaxy formation model~\citep{Weinberger:2017,Pillepich:2018}, which was developed from the original Illustris model/simulation~\citep{Vogelsberger:2014,Genel:2014}. The model includes gas cooling, both primordial and from metal lines, where nine chemical elements are tracked; star formation and stellar evolution, which includes kinectic-wind feedback from Type-II supernovae, and metal enrichment from supernovae (Types Ia and II) and asymptotic giant branch stars~\citep{Pillepich:2018}; massive black hole growth, carrying two modes of feedback: thermal injection for high accretion rates, and kinetic winds at low accretion rates~\citep{Weinberger:2017}; and an idealized consideration of magnetic fields~\citep{Pakmor:2013}. The main observational constraints used to calibrate the model included the $z=0$ galaxy stellar mass function, the $z=0$ stellar-to-halo mass relation, and cosmic star formation rate density history~\citep[for details, see][]{Pillepich:2018}. 

In this paper, we use the TNG100 simulation, first presented in a series of five papers~\citep{Marinacci:2018,Naiman:2018,Nelson:2018,Pillepich:2018b,Springel:2018}, which has been made publicly available~\citep{Nelson:2019}. The periodic box length of TNG100 is $75\hmpc\approx 110$ comoving Mpc, within which $1820^3$ dark-matter particles and $1820^3$ initial gas cells are evolved. This gives a mass resolution of $7.5\times 10^6 {\rm M}_\odot$ for dark matter and $\sim 1.4\times 10^6 {\rm M}_\odot$ for baryons. The smallest resolvable length for gas in galaxy centres is 190~pc. The simulation has been post-processed to decompose atomic gas into its atomic and molecular phases; we follow the methodology of \citet{Stevens:2019}, which is based largely on \citet{Diemer:2018}. 

TNG100 galaxy properties in this paper follow the `inherent' definition of~\citet{Stevens:2019}. This means isolating particles/cells bound to (sub)structures according to {\sc Subfind}~\citep{Springel:2001,Dolag:2009} and further making an aperture cut to remove the near-isothermal `intrahalo' component~\citep{Stevens:2014}.%
\footnote{We note a previously unreported error in the application of this aperture in the results of \citet{Stevens:2019}, which meant the radius always hit its upper limit of $R_{\rm 200c}$.  We have corrected for this in this work.  The effect is negligible for our results and that of \citet{Stevens:2019}.}
All \HI/\Hmol~properties assume the \citet{Gnedin:2014} prescription, as described in \citet{Stevens:2019}. The primary factors in this prescription are the gas density, local UV flux, and local dust content. Critically, it is assumed that UV is generated from sites of star formation, with 90\% of that UV absorbed in the star-forming cell; the other 10\% is propagated through the halo, treating it as a transparent medium~\citep{Diemer:2018}. A lower limit on the UV flux in any cell is set to the assumed cosmic UV background~\citep{Faucher:2009}. The dust fraction of each cell is assumed directly proportional to its metallicity, in line with \citet{Lagos:2015}.  A comparison of the performance of this prescription with several others has been presented alongside many \HI- and H$_2$-related properties of TNG100 galaxies in a series of recent papers~\citep{Diemer:2018,Diemer:2019,Stevens:2019,Stevens:2019b}. 

\romeel{
\subsection{Comparison of Models}
\label{sec:aperture}

For clarity, we briefly summarise the key differences between our cosmological simulations, in terms of measuring the cold gas properties.  All employ the \citet{Rahmati:2013} prescription to determine the self-shielded gas, and all employ some variant of the \citet{Krumholz:2011} model to compute the H$_2$ fraction.  \simba\ does these on-the-fly, while the others apply these in post-processing, but since their impact on the dynamics should be relatively minimal, there should not be large variations because of this.

Assigning gas into galaxies has more variations between models.  In \simba, all \HI\ and H$_2$ in halos (which  is essentially all cold gas overall) is assigned to the most dynamically important galaxy (i.e. maximum in $M_{\rm gal}/r^2$ where $r$ is the distance to the galaxy's center of mass).  While we identify the galaxies using a 6DFOF and use this to compute properties such as the SFR and $M_*$, only the locations are used for the cold gas contents, with all cold gas being assigned via proximity.  In the case of H$_2$, this is essentially equivalent to just assigning the H$_2$ to its own galaxy. The \HI, however, can extend well beyond the star-forming region of a galaxy, so this procedure can yield large differences in \HI\ content. The choice for \simba\ is motivated by our comparisons to Arecibo data (see next section) which has arcminute-scale resolution, and thus is unlikely to be resolving individual galaxies' ISM.  

EAGLE, \romeelnew{also motivated by matching Arecibo's beam~\citep{Bahe:2016}}, uses a fixed 70~kpc aperture to compute both \HI\ and H$_2$ contents, within which only gas that is gravitationally bound to a given galaxy is attached to it.}
~\adam{\TNG, meanwhile, uses the variable `BaryMP' aperture of \citet{Stevens:2014}.  
For a fixed stellar mass of $10^9\msolar$ at $z=0$, the aperture radius comes out as $20\pm10$ kpc (rounded 1$\sigma$ range).  For $M_*=10^{11}\msolar$, this rises to $70\pm30$\,kpc.  For the same respective stellar masses at $z=2$, aperture radii are $15\pm5$ and $40\pm10$ physical kpc.
A similar exercise at fixed $\mhi$~of $10^8$ and $10^{10}\msolar$ yields aperture radii of $16\pm4$ and $50\pm9$ kpc, respectively, at $z=0$.
To first order, the aperture scales with the virial radius of each system, where typically $R_{\rm BaryMP}/R_{\rm vir} \simeq 0.25\pm0.10$.}
~\romeel{This means that in EAGLE, the largest galaxies may be missing gas that is at large radii compared to what is in TNG.  This is not expected to be significant for H$_2$ where the gas is mostly confined to the dense ISM, but may be important in the case of \HI\ that can extend quite far out.  On the flip side, in small galaxies EAGLE may include \HI\ that TNG would not include, but in such systems the \HI\ is not expected to extend very far out.  \simba\ does not use aperture masses, but in principle can assign \HI\ from anywhere in the halo, so it is not straightforward to directly compare to aperture masses, but in general \simba\ includes \HI\ from at least as far out as TNG.  We emphasize, however, that in all simualtions most of the \HI\ is located relatively close to galaxies.

We can compare these radii to the observed \HI\ size--mass relation, $R_1 = \sqrt{\mhi/12.88}$ where $R_1$ in pc is the radius at which the surface density drops to 1~M$_\odot$~pc$^{-2}$, and $\mhi$ is in solar masses \citep{Stevens:2019b}}\romeel{.  This gives, for $\log\ \mhi=[8,9,10,11]$, $R_1 = [2.8,8.8,28,88]$~kpc.  Thus generally  the TNG aperture extends out to at least this radius, while EAGLE's aperture is not expected to miss significant \HI\ except for $\mhi\ga 10^{10} M_\odot$.  \citet{Bahe:2016} showed that EAGLE's choice of 30~kpc can bias the resulting $\mhi$ by up to $\sim 20\%$, which is not negligible but does not impact the conclusions of this paper.

Given that the \HI\ distribution is fairly extended and thus more subject to choices regarding aperture or assignment, we thus emphasise that \HI\ comparisons between models and to ALFALFA data should be regarded as preliminary; \romeelnew{this could either overestimate or underestimate the \HI\ content depending on the assumed aperture relative to the observed beam.} Nonetheless, the differences between models are significantly larger than the expected systematics associated with apertures. Upcoming observations with sensitive radio interferometers such as MeerKAT and ASKAP will avoid blending issues~\citep{Elson:2019}, but may require even more care when considering \HI\ apertures in simulation comparisons; we leave a fuller investigation of this for future work.

For H$_2$, we have a somewhat different issue, because the beam size for our main comparison sample xCOLD~GASS, based on IRAM-30m data, is a factor of $\sim$5 smaller than Arecibo, with typical apertures of $\sim 10-20$~kpc.  Aperture corrections are applied to these observations, as detailed in \S2.4 of \citet{Saintonge:2017} as well as in \citet{Saintonge:2012}; these indicate that the observations appear to capture the majority of H$_2$ in these systems.
However, if one applies similarly small apertures to the simulations, one can get non-trivially different results, particularly for massive gas-rich galaxies that are quite large in extent.  The underlying problem is probably that all the simulations assume that H$_2$ is allowed to form in gas with $n_H\ga 0.1-0.2$~cm$^{-3}$, whereas in reality H$_2$ actually forms only at much higher densities ($n_H\ga 100$~cm$^{-3}$).  The choice in these simulations is driven by the numerical resolution, which does not enable the internal structure of the ISM to be fully represented.  The result is that H$_2$ can form in more extended gas in simulations than in real data.  Unfortunately, this is an intrinsic limitation even in these state-of-the-art models, and suggests that better sub-resolution ISM models are required to properly represent the distribution of molecular gas in galaxies.

Ultimately, the models and observations are both including the vast majority of H$_2$ in the galaxies.  However, the spatial distribution in the case of simulations is more extended owing to resolution limitations.   \citet{Diemer:2019} demonstrated this explicitly, showing that the molecular gas distribution in \TNG\ is more extended than observed.
Hence if we were to create mock observations with the same aperture restrictions, we may obtain substantially smaller H$_2$ contents for the most massive gas-rich galaxies.
But given the limitations in way that H$_2$ is modeled in the ISM, this would not be a meaningful way to conduct such a comparison.  We thus will conduct our comparisons with the apertures as described above, with the caveat that we are comparing only the {\it global} molecular gas contents, without creating mock CO images tailored to individual surveys.
}

\adam{In truth, none of the decisions surrounding apertures in this paper are necessarily faithful representations of the scale on which gas (or stellar) properties of galaxies are measured observationally.  
For \HI~in \TNG~galaxies, this has been explored by \citet{Stevens:2019} and \citet{Diemer:2019}.
In the former, alongside the inherent \TNG~properties, mock-observed properties that were catered to specific \HI~surveys were presented, namely for ALFALFA and xGASS.
Mock and inherent \HI~measurements differed most for satellite galaxies, primarily due to the much higher probability of `confusion' (where \HI~in the mock beam comes from sources other than the galaxy of interest).
A similar mocking procedure for H$_2$ will be presented for TNG100 in Stevens et al.~(in prep.). Finally, we note that aperture choices can also be important for stellar mass measures, particularly for massive galaxies~\citep[e.g.][]{Bernardi:2017}, so these issues are not limited to gas measures.
}

\subsection{xGASS and xCOLD GASS} 

The twin surveys xGASS and xCOLD GASS were specifically designed to provide a robust and complete census of \HI\ and CO across the $z\sim0$ Universe for purposes such as comparing and constraining large-scale simulations. Unlike most previous gas surveys, they are neither flux-limited nor the result of complex selection functions; the galaxies are selected from SDSS based on stellar mass and redshift alone, and are therefore representative of the local galaxy population. 

The \HI\ component of this programme, xGASS, provides 21-cm measurements for 1179 galaxies with $10^9<M_\ast/{\rm M}_\odot <10^{11.5}$ and $0.01<z<0.05$ \citep{Catinella:2010,Catinella:2013,Catinella:2018}. The sample is extracted from the overlap of the SDSS spectroscopic and GALEX imaging surveys, and builds on ALFALFA \HI-blind survey \citep{Giovanelli:2005}, which provides the measurements for the most gas-rich systems in the sample. 

In addition to the sample selection, the observing strategy employed by xGASS is key to making it fit for purpose here. The observations were designed to provide uniform sensitivity in terms of atomic gas mass fraction, $\fneut=\mhi/M_*$, which allows to compare gas contents with global galaxy properties over a wide range of stellar masses. Apart from the galaxies already detected in \HI\ by ALFALFA, each xGASS galaxy was observed with Arecibo until either a detection of the \HI\ line was obtained or until sensitivity to a gas mass fraction of 2-10\% (depending on stellar mass) was reached. For \HI\ non-detections, 5$\sigma$ upper limits have been computed. 

The xCOLD GASS survey targeted 532 galaxies from xGASS with the IRAM-30m telescope to measure their total molecular gas masses via CO(1-0) line emission \citep{Saintonge:2011,Saintonge:2017}. As for the \HI, the strategy was to observe each galaxy until either a detection of the CO(1-0) line was obtained or until sensitivity to $\fmol=M_{\rm H_2}/M_*=2$\% was reached, allowing stringent upper limits to be placed on the most gas-poor galaxies. The CO line luminosities are corrected for small aperture effects \citep{Saintonge:2012}, and then converted into molecular gass masses using the CO-to-H$_2$ conversion factor $\alphaCO$ of \citet{Accurso:2017}.

\citet{Accurso:2017} characterised $\alpha_{\rm CO}$ using [CII] data from {\it Herschel} and CO(1-0) data from xCOLD~GASS, guided by photo-dissociation region modeling. They found that $\alpha_{\rm CO}$ depends quite strongly on the galaxy's metallicity, and weakly on its deviation from the star-forming main sequence:
\begin{equation}
    \log\ \alpha_{\rm CO}(\pm 0.165) = 14.752 - 1.623(12+\log({\rm O/H})) + 0.062\log\Delta_{MS},
\end{equation}
where $12+\log({\rm O/H})$ is the oxygen abundance in conventional notation, and $\Delta_{MS}$ is the ratio of the SFR in the galaxy to that in a galaxy on the main sequence at the same $M_*$.  Using $12+\log({\rm O/H})_\odot=8.69$~\citep{Asplund:2009}, we can rewrite this as:
\begin{equation}\label{eq:XCO_accurso}
    \alpha_{\rm CO} = \frac{4.47^{+2.06}_{-1.41}}{Z_{\rm H_2}^{1.623}\Delta_{MS}^{0.062}}\;\;\aCOunits,
\end{equation}
We will show a comparison of this method for determining $\alphaCO$ versus the simulation-based method we employ from \citet{narayanan12a} in \S\ref{sec:alphaCO}.

The \HI\ and CO observations are complemented by additional data products. Stellar masses are retrieved from the MPA-JHU catalog, and stellar mass surface densities were calculated as $\mu_*=M_*/(2 \pi r^2_{50,z})$, where $r_{50,z}$ is the radius encompassing 50\% of the $z$-band flux, in kpc. Star formation rates were calculated using a combination of UV and IR photometry and an optical SED fitting method \citep{janowiecki17}. 



\section{Results}\label{sec:results}

\subsection{Stellar and SFR properties}

\begin{figure}
	\includegraphics[width=0.48\textwidth]{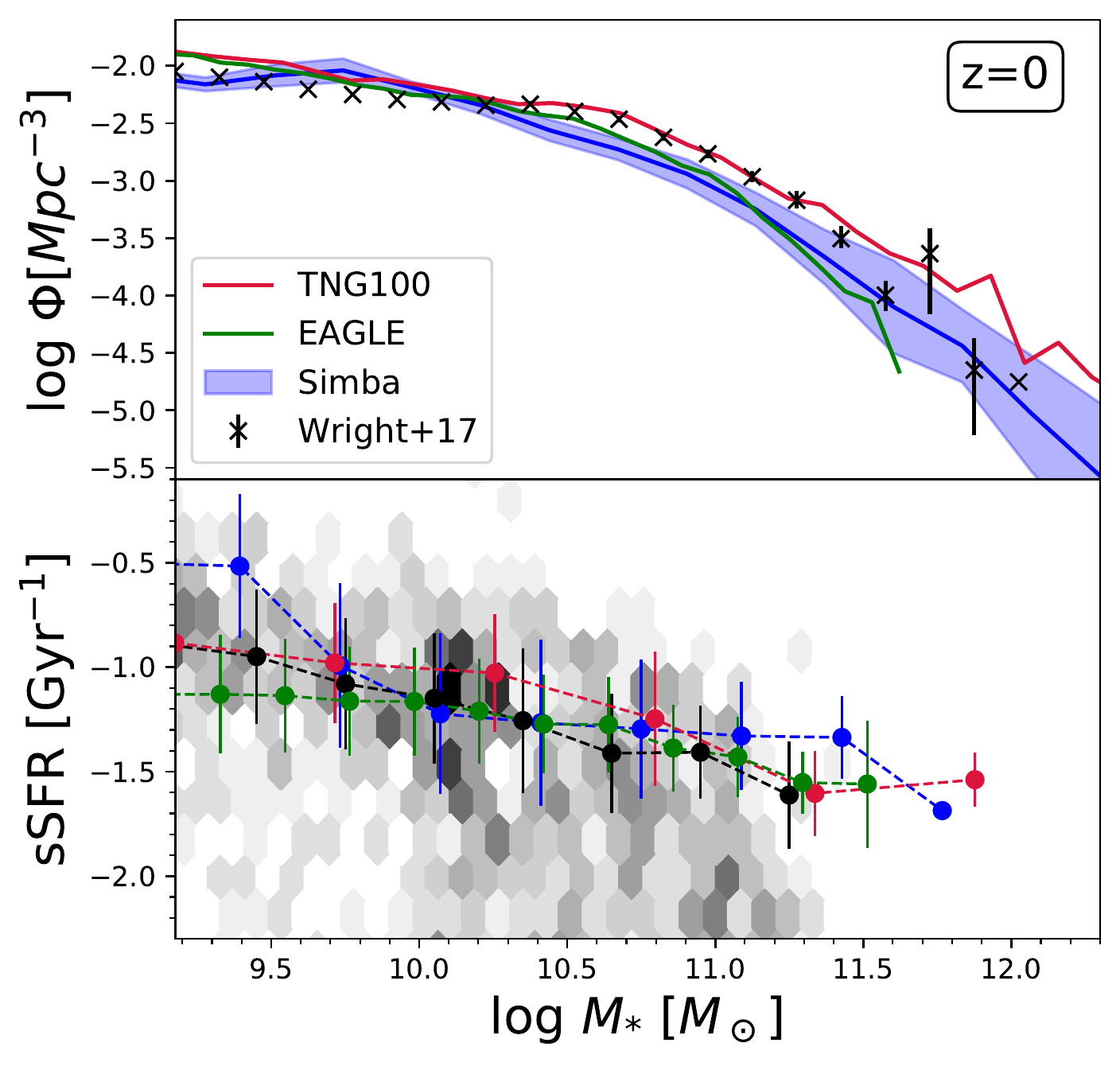}
	\vskip-0.1in
    \caption{The galaxy stellar mass function (top panel) and specific star formation rate vs. stellar mass (bottom panel) at $z=0$ for \simba\ (blue), \TNG\ (red), and EAGLE (green) compared with observations.  The blue shading for \simba\ shows the variance over 8 simulation sub-octants.  For the GSMF, representative observations are shown from \citet{Wright:2017}.  For the bottom panel, the hexbin shows the xGASS sample.  Running medians are shown for galaxies with sSFR$>10^{-1.8}$~Gyr$^{-1}$; dashed black line shows xGASS, while the red, green, and blue lines \TNG, EAGLE, and \simba, respectively.  Error bars show the $1\sigma$ spread around the median.  All models show good agreement with these stellar-based measures, with some minor discrepancies.}
    \label{fig:ms_sfr}
\end{figure}

To set the stage for our gas comparisons, we first present and compare stellar mass and star formation rate properties in our three simulations.  In particular, we look at the galaxy stellar mass function and the star formation rate--stellar mass relation (``main sequence''). These have been presented in other papers~\citep{Furlong:2015,Pillepich:2018,Dave:2019}, but here we provide a direct comparison to each other as well as to observations. For the main sequence, we will compare to the galaxies in the xGASS sample in order to ensure there are no clear systematic differences in the sample properties.

Fig.~\ref{fig:ms_sfr}, top panel, shows the $z=0$ galaxy stellar mass function (GSMF) in \simba\ (blue), \TNG\ (red), EAGLE (green).  The blue shaded region shows the standard deviation computed over 8 sub-octants of the \simba\ volume, as an estimate of cosmic variance; since the other simulations have similar volumes, they likely have comparable variance.  The observed $z\approx 0$ GSMF is shown from \citet{Wright:2017} as the black data points, from the Galaxy And Mass Assembly (GAMA) survey.  We do not show EAGLE-Recal here to avoid clutter, but it was specifically recalibrated to match the GSMF, so agrees as well as the main EAGLE simulation.

All simulations provide a good match to the observed $z=0$ GSMF.  In part, this is by construction, as the star formation and feedback recipes in each have been tuned at some level to reproduce this key demographic. They agree quite well below $M^\star$, but there some minor differences at the massive end.  EAGLE and \simba\ slightly undercut the knee of the GSMF, while \TNG\ matches the knee very well but may overproduce the massive end. This highlights the continued difficulty that simulations have in reproducing the sharpness of the exponential cutoff in the $z=0$ GSMF~\citep[e.g.][]{Dave:2016}. Note that the massive end is relatively uncertain owing to aperture effects and potentially initial mass function variations~\citep[e.g.][]{Bernardi:2018}.   Despite these small variations, in general all these simulations reproduce the observed GSMF quite well, within plausible systematic uncertainties.

Fig.~\ref{fig:ms_sfr}, bottom panel, shows specific star formation rate (sSFR=SFR/$M_*$) versus stellar mass for \simba\ (blue), \TNG\ (red), and EAGLE (green).  Here, we show a running median for each simulation for star-forming galaxies defined as having sSFR$>10^{-1.8}$Gyr$^{-1}$, following \citet{Dave:2019}.  The grey hexbins show the mass-selected xGASS sample, and the black points and line show a similarly selected running median of the xGASS star-forming galaxies.  All have errorbars representing the $1\sigma$ scatter around the median.  We focus on a comparison of the star-forming sample since we are primarily interested in gas-rich galaxies in this work; as an aside, there are larger differences in the quenched galaxy fractions among these various models.

All models produce a mildly declining relation of sSFR vs. $M_*$ in non-quenched galaxies that is in good agreement with the xGASS sample, as well as with each other.  There are mild differences in the detailed shape of the curves, such as \simba\ producing high sSFR values at low masses,  \TNG\ having slightly higher sSFR values around $M^\star$, and EAGLE potentially slightly low at low masses.  The hexbinned xGASS sample shows a turndown in sSFR at high masses, which would also be evident in the simulations' running medians if we were to include sSFR$<10^{-1.8}$Gyr$^{-1}$ galaxies.  The $1\sigma$ scatter is typically in the range of $\sim 0.3$~dex in all models, which is comparable to that seen in the observations~\citep[e.g.][]{Kurczynski:2016,Catinella:2018}.

Overall, we confirm that \simba, \TNG, and EAGLE all produce stellar mass functions and star-forming main sequences that are in good agreement with observations, and with each other.  This is an important check, which sets the baseline for comparisons among their cold gas properties in relation to their $M_*$ and SFR.  It is also a non-trivial success for these models, which has only been achieved in the last few years among hydrodynamic galaxy formation simulations.  Nonetheless, we will see that the differences among gas properties in these simulations are significantly larger than those seen in stellar properties.

\subsection{\HI\ Mass Function}
\label{sec:HIMF}

\begin{figure}
	\includegraphics[width=0.45\textwidth]{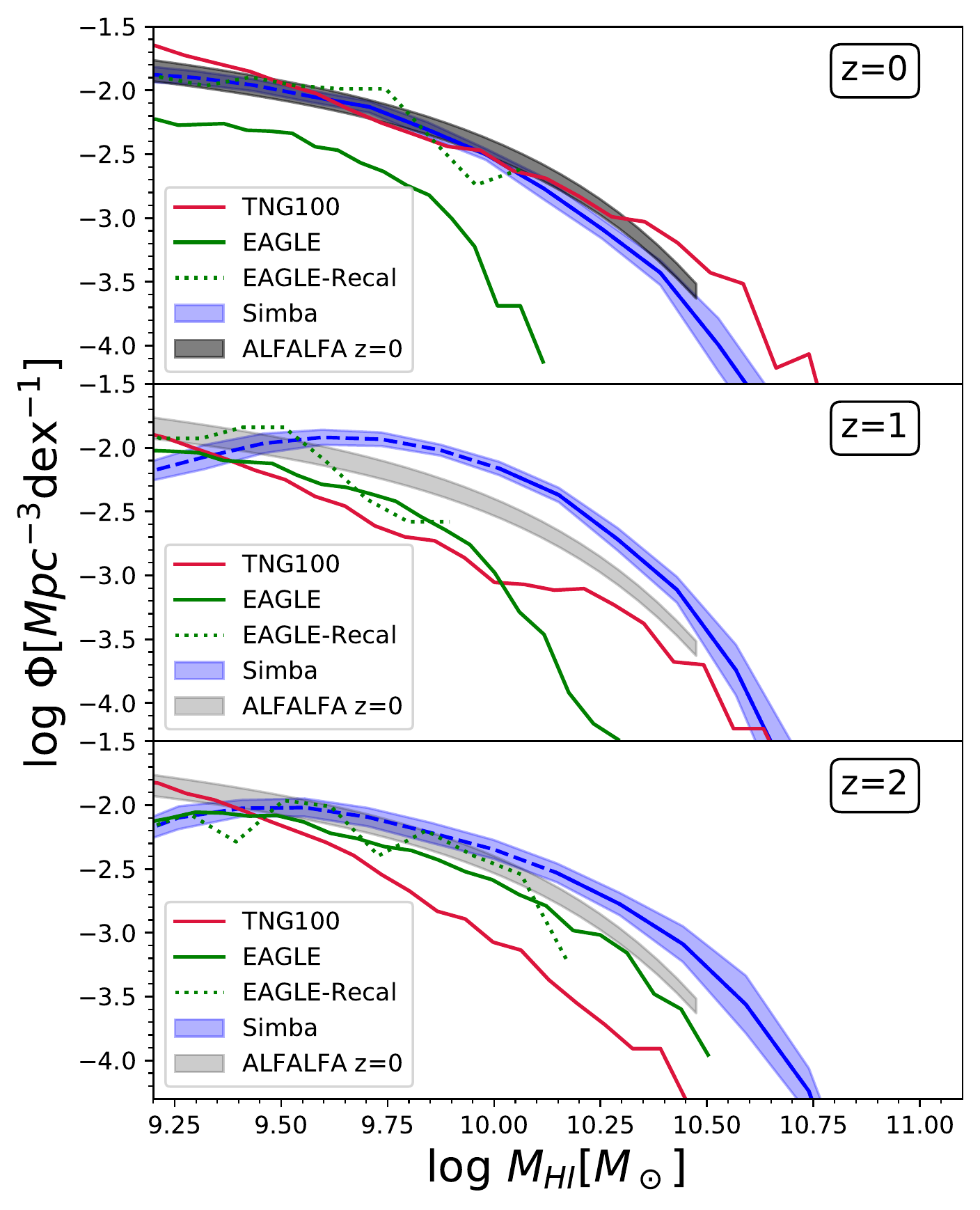}
	\vskip-0.1in
    \caption{The \HI\ mass function in \simba\ (blue), \TNG\ (red), and EAGLE (green), at $z=0,1,2$ (top to bottom).  The HIMF from the ALFALFA survey~\citep{Jones:2018} at $z\approx 0$ is shown as the grey shading, and these are reproduced at higher $z$ in lighter shading for reference.  \TNG\ and \simba\ shows much more \HI\ than EAGLE, in better agreement with observations, while EAGLE and \simba\ show an increasing HIMF at earlier epochs while \TNG\ shows a dropping HIMF.}
    \label{fig:HIMF}
\end{figure}

We now examine properties of the neutral gas, starting with the \HI\ mass function (HIMF) and its redshift evolution in our three simulations. The \HI\ mass function is straightforwardly measurable from 21-cm emission down to quite low masses, albeit currently only at low redshifts owing to the sensitivity limits of radio telescopes. Hence the HIMF has relatively few concerns regarding selection effects or other observational systematics, modulo confusion issues given the large beam sizes of single-dish surveys~\citep[e.g.][]{Elson:2019}.  That said, there are some non-trivial modeling systematics regarding assumptions about self-shielding and the separation of atomic and molecular hydrogen, although since the neutral gas content tends to be dominated by atomic gas particularly in small systems, this is less of an uncertainty for \HI\ as for H$_2$.  For these reasons, the \HI\ mass function is among the more robust constraints on the cold gas content in simulated galaxies.

Fig.~\ref{fig:HIMF} shows the \HI\ mass function at $z=0,1,2$ (top to bottom) for \simba\ (blue), \TNG\ (red), EAGLE (green solid), and EAGLE-Recal (green dotted).  The blue shaded region shows the estimated cosmic variance computed over 8 sub-octants of the \simba\ volume. At $z=0$, we show the observed HIMF from the ALFALFA survey~\citep{Jones:2018} as the dark grey band; we repeat this in the $z=1,2$ panels with lighter shading as a reference point to gauge the amount of evolution in models, but current observations of the HIMF are limited to low redshifts so comparisons to data should only be done at $z=0$.  
\romeel{Finally, for \simba, we will show in \S\ref{sec:simba_agn} that, owing to its relatively low resolution compared to \TNG\ and EAGLE, it suffers from incompleteness at the low-$\mhi$ end.  To denote this, we have shown the portion of the HIMF that is potentially compromised by resolution effects using a dashed blue line.}

At $z=0$, all simulations show a fairly flat low-mass slope, and a turnover at high masses; broadly, this is similar to that seen in ALFALFA. To quantify this, we fit a Schechter function to each simulated HIMF for galaxies with $\mhi>10^9 {\rm M}_\odot$.  Given the limited dynamic range, we fix the low-mass slope to the ALFALFA value of $-1.25$; leaving the slope free gives values consistent with this, but with significantly larger uncertainties on all the parameters.  We find that the best-fit characteristic \HI\ mass $\mhi^\star$ varies significantly between models: For \TNG\ it is $\mhi^\star=10^{10.26}{\rm M}_\odot$, for \simba\ it is $\mhi^\star=10^{10.07}{\rm M}_\odot$, while for EAGLE it is $\mhi^\star=10^{9.67}{\rm M}_\odot$. For comparison, ALFALFA finds $\mhi^\star=10^{9.94}{\rm M}_\odot$~\citep{Jones:2018}.  This confirms the visual impression that \TNG\ \romeel{while generally matching the HIMF as seen in \citet{Diemer:2019}, mildly over-predicts the HIMF at the high-mass end; we note that \citet{Diemer:2019} supplemented TNG-100 with TNG300 and showed that it provided a better match to the high-mass end.}~ Meanwhile the main EAGLE volume strongly under-predicts the HIMF, \romeel{while the higher-resolution $25\hmpc$ EAGLE-Recal simulation produces a significantly higher HIMF.}  \simba\ provides a very good match to ALFALFA observations in both shape and amplitude.  Note that none of these models have been tuned to reproduce the HIMF.

As noted by \citet{Crain:2017} and seen in Fig.~\ref{fig:HIMF}, EAGLE-Recal produces significantly more \HI\ in galaxies, although its volume is too small to probe the high-mass turnover discrepancy. Hence in EAGLE the \HI\ content appears to be fairly resolution-dependent, which we speculate is likely a consequence of EAGLE's subgrid implementation of feedback (intentionally) not incorporating mechanisms to mitigate against resolution sensitivity (as is the case for \simba\ and \TNG). As noted by \citet{Bahe:2016}, the thermal energy injected into the ISM by feedback events in EAGLE scales linearly with the baryon particle mass, and at the standard resolution of EAGLE individual heating events can temporarily create $\sim$kpc-scale `holes' in the cold gas distribution.  \romeel{Assuming higher resolution simulations produce more robust results, the EAGLE-Recal results suggest that the EAGLE feedback model is capable of well reproducing the HIMF.}

The HIMF resolution convergence was generally good in the case of \simba's predecessor, \mufasa~\citep{Dave:2017a}.  Because the star formation feedback is quite similar in \simba, we expect this would also be the case in \simba\ either, and we will demonstrate this in \S\ref{sec:simba_agn}.  \romeel{Thus to avoid clutter, we choose not to show different resolution versions of these simulations, though we show EAGLE vs. EAGLE-Recal.
For \TNG, \citet{Diemer:2019} showed the level of convergence in the cold gas mass functions between TNG100 and TNG300\footnote{\romeelnew{For a full exploration of resolution convergence in \TNG, see \tt http://www.benediktdiemer.com/data/hi-h2-in-illustris/}}, and found that they were generally in agreement with each other in their overlapping resolved mass ranges, with TNG300 being slightly lower.  This suggests that models that use kinetic decoupled winds may fare somewhat better in resolution convergence than thermal feedback models.  Note, however, that all these comparisons are subject to aperture effects that could cause more significant changes than resolution (\S\ref{sec:aperture}).
}

Even stronger differences between the simulations are seen when examining redshift evolution.  \simba\ and EAGLE both have strong redshift evolution, in the sense that the HIMF shifts to higher $\mhi$ at higher redshifts.  This occurs owing to the higher inflow rates at higher redshifts, which in the quasi self-regulated scenario for galaxy growth results in higher gas contents~\citep[see e.g. Fig. 5 of][]{Crain:2017}. Quantified purely in terms of characteristic mass evolution (fixing the low-mass slope) from $z=0\to 2$, $\mhi^\star$ increases by $\times 2.5$ and $\times 2$ in \simba\ and EAGLE, respectively.  Interestingly, EAGLE-Recal does not show as much evolution as EAGLE, and is consistent with the larger volume at $z=1,2$.  Meanwhile, \TNG\ shows a {\it reduction} in $\mhi^\star$ by $\sim\times 2$ between $z=0\to 2$; it is not immediately evident why \TNG\ shows this behaviour.  In \TNG\ and EAGLE, the low-mass slope becomes steeper at high-$z$.  \simba\ shows a turnover at low masses ($\mhi\la 10^9 {\rm M}_\odot$), but this owes largely to numerical resolution, as we show in \S\ref{sec:simba_agn}.  At higher redshifts, the higher $\mhi/M_*$ ratios together with the fixed galaxy mass threshold of $M_*=5.8\times 10^8 {\rm M}_\odot$ combine to result in significant incompleteness at $\mhi\la 10^{9.5}{\rm M}_\odot$ by $z\sim 2$.  Modulo these caveats that mostly impact the low-$\mhi$ end, it is clear that measurements of the HIMF even out to $z\sim 1$, as is planned with MeerKAT's LADUMA survey~\citep{Blyth:2016}, could provide qualitative discrimination between current galaxy formation models.

\subsection{H$_2$ Mass Function}

\begin{figure}
	\includegraphics[width=0.45\textwidth]{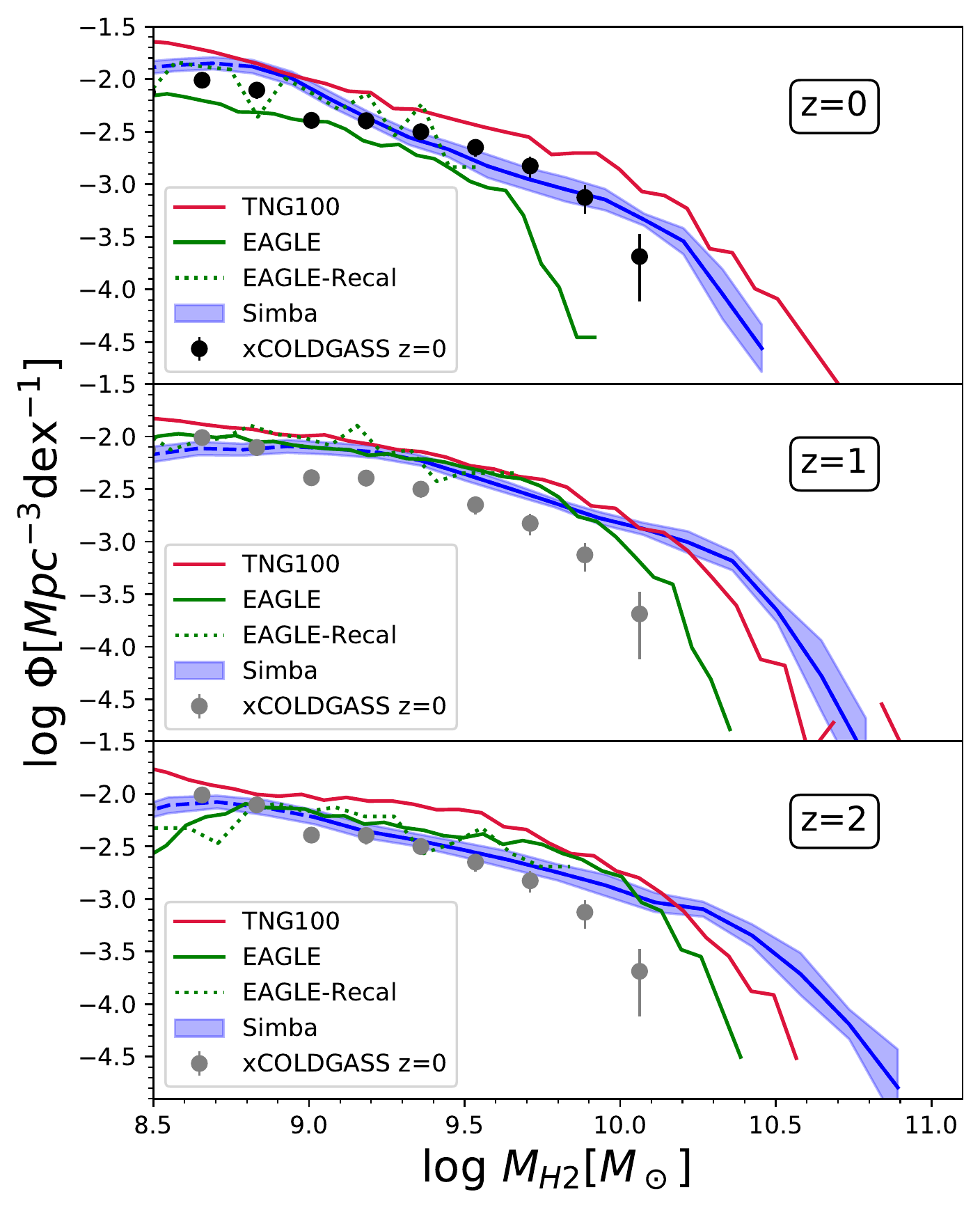}
	\vskip-0.1in
    \caption{The H$_2$ mass function in \simba\ (blue), \TNG\ (red), and EAGLE (green), at $z=0,1,2$ (top to bottom).  Observations from xCOLD~GASS~\citep{Fletcher:2020} at $z\approx 0$ are shown as the black points, and reproduced in grey at higher $z$ for reference.  \TNG\ and \simba\ yield significantly larger molecular masses than EAGLE.  \simba\ and EAGLE show an increased H2MF at the high-mass end at $z=2$, while \TNG's is lower. }
    \label{fig:H2MF}
\end{figure}

Stars form from molecular gas, so the molecular gas content provides a connection to the growth rate of galaxies, particularly in star-forming systems where the merger growth rate is sub-dominant~\citep[e.g.][]{Hirschmann:2014}. This has been explored in previous works including \citet{Lagos:2015} for EAGLE and \citet{Dave:2019} for \simba.  However, as emphasised in analytic ``equilibrium'' or ``bathtub'' models of galaxy evolution~\citep[e.g.][]{Finlator:2008,Bouche:2010,Dave:2012,Lilly:2013}, the molecular gas content does not govern the global stellar mass assembly history, but rather represents an evolving balance between gas supply and gas consumption.  For a given gas supply into the ISM, if the star formation efficiency (SFE=SFR$/M_{\rm H_2}$) is high, then the gas reservoir will be low, and vice versa, though the time-averaged number of stars formed will not be altered. Meanwhile, the star formation history of a galaxy over cosmological timescales is set primarily by the net gas supply rate (inflows minus outflows), and is globally independent of SFE for reasonable choices~\citep{Katz:1996,Schaye:2015}. The molecular gas reservoir thus represents a way to characterise this SFE.  In observational work, this is often presented as measures of its its inverse quantity, the molecular gas depletion time.

In cosmological simulations, the SFE is an input parameter.  Typically, it is tuned to approximately reproduce the \citet{Kennicutt:1998} relation in star-forming galaxies~\citep[e.g.][]{Springel:2003}, which for instance has been checked in \simba~\citep{Appleby:2019}. In practice, however, the SFE parameter is applied in \simba\ and \TNG\ via a volumetric \citet{Schmidt:1959} law, which is connected to the \citet{Kennicutt:1998} surface density relation via galactic structure; EAGLE uses a scheme based on the local pressure which results in the \citet{Kennicutt:1998} relation by construction in the case of vertical hydrostatic equilibrium.  In all cases, the molecular gas content in simulations is thus also sensitive to the distribution of molecular gas within galaxies.  Furthermore, as mentioned in \S\ref{sec:simba}, \simba's star formation prescription uses a molecular gas-based \citet{Schmidt:1959} law, while that in \TNG\ and EAGLE use the total gas, which could result in further differences in the internal structure of star-forming gas.  For these reasons, the H$_2$ mass function provides insights into the differences between current star formation prescriptions, particularly among simulations that well reproduce the observed growth histories of the galaxy population.

Fig.~\ref{fig:H2MF} shows the H$_2$ mass function (H2MF) at $z=0,1,2$ (top to bottom) for \simba\ (blue), \TNG\ (red), EAGLE (green solid), and EAGLE-Recal (green dotted), with a blue shading on \simba\ for the estimated cosmic variance as before.  Observations at $z\approx 0$ are shown as the black points from the xCOLD GASS survey \citep{Fletcher:2020}, and reproduced at other redshifts in grey for reference.  At higher redshifts, we will compare more directly to CO luminosity functions in \S\ref{sec:COLF}.

As with the \HI\ mass function, the general Schechter function shape is found in all simulations, but there are substantive differences between model predictions.  At $z=0$, \simba\ and \TNG\ produce nearly identical H2MFs, while EAGLE's is much lower.  The differences are particularly dramatic at the high-mass end.  A Schechter fit fixing the observed faint-end slope at $-1.33$~\citep{Saintonge:2017} gives $M^\star_{\rm H_2}=10^{9.98}$ for \simba, $M^\star_{\rm H_2}=10^{10.07}$ for \TNG, but $M^\star_{\rm H_2}=10^{9.21}$ for EAGLE.  Comparing to the observed $M^\star_{\rm H_2}=10^{9.68}$ from xCOLD~GASS highlights the discrepancies between all these models and the observations, with \simba\ and \TNG\ overproducing the H2MF at the massive end, while EAGLE under-predicts it. \romeel{We reiterate here that the comparisons at the massive end are potentially subject to uncertainties regarding aperture effects, given that the simulations' apertures are generally significantly larger than in the observations, as discussed in \S\ref{sec:aperture}.}

At the low-mass end, \simba\ shows a dip that owes to limited numerical resolution; \romeel{we denote the low-mass portion of the H2MF that is subject to resolution effects via the dashed blue line}~(see \S\ref{sec:simba_agn}). However, it agrees around the knee of the H2MF, while \TNG\ over-predicts the H2MF somewhat at all masses.  These differences may be subject to significant systematics, not the least of which is the assumed CO-to-H$_2$ conversion factor used to determine $M_{\rm H_2}$ from the observations, as we will explore in \S\ref{sec:alphaCO}.

The shifts to higher redshifts again shows a similar pattern as the HIMF:  The H2MF is clearly increasing to higher redshifts in \simba\ and EAGLE, but mostly unevolving in \TNG. For EAGLE, $M^\star_{\rm H_2}$ increases by $\times 4$ between $z=0$ and $z=2$, for \simba\ it is $\times 2.5$, while there is no clear increase for \TNG.  Comparing to $M^\star_{\rm HI}$ evolution, we see then that EAGLE and \TNG\ both yield a greater increase (or less decrease) in $M^\star_{\rm H_2}$ from $z=0$ to $z=2$, while for \simba\ the increase is similar in both \HI\ and H$_2$; for \simba, this is reflected in the similarity of the evolution of $\Omega_{\rm HI}$ vs. $\Omega_{\rm H_2}$ as noted in \citet{Dave:2019}.

Overall, the H2MF shows strong differences between models both at $z=0$ and in terms of evolution to higher redshifts.  This highlights the potential for molecular gas mass measurements to be a key discriminator between models.  We will discuss the differences in input physics that may be causing these variations in \S\ref{sec:discussion}, but it is interesting that none of the models reproduce the $z\approx 0$ H2MF ``out of the box''.  We note that \citet{Lagos:2015} found better agreement between EAGLE's H2MF and observations from \citet{Keres:2003} when converting their data to H$_2$ masses assuming a constant $\alphaCO=2 \aCOunits$, but this value is low compared to the canonical value for Milky Way-like galaxies of $\alphaCO\approx 4.5$, \romeel{\citet{Diemer:2019} found better agreement between \TNG\ and the H2MF inferred by \citet{Obreschkow:2009}, but the recent xCOLD~GASS determination is somewhat lower at low masses, resulting in more of an apparent disagreement.  
At the massive end, we are using a significantly larger aperture than \citet{Diemer:2019} in order to capture all the molecular gas and compare the global H$_2$ content, but this increases the H$_2$ content of these massive systems into poorer agreement.  \citet{Popping:2019} likewise found that apertures can have a significant impact on this comparison, and only including mass within a 3.5" aperture resulted in significantly better agreement.  As discussed in \S\ref{sec:aperture}, it is not entirely clear which way of doing the comparison is more correct.  But what this does indicate is that
}
systematic uncertainties in both observing and modeling the H2MF may be substantial.  Among the most crucial of these is the conversion factor between observed CO luminosity and the H$_2$ mass.  We examine this issue next, and use this to bring our comparisons into the observational plane of CO(1-0) luminosities.

\subsection{The H$_2$-to-CO conversion factor}
\label{sec:alphaCO}

Observationally, the molecular hydrogen content is most commonly traced via the CO luminosity.  Since simulations most directly model molecular gas, we need to convert the molecular gas mass into a CO luminosity.
This conversion factor, known as $\alpha_{\rm CO}\equiv M_{\rm H_2}/L_{\rm CO}$, is the subject of much debate~\citep[see the review by][]{Bolatto:2013}. It is clear that $\alpha_{\rm CO}$ depends on metallicity as well as the local strength of H$_2$ dissociating radiation, which in turn depends on quantities such as the local star formation rate and shielding column density.  For fairly massive galaxies with close to solar metallicity, it is observed that Milky Way-like galaxies have $\alpha_{\rm CO}\approx 4.5 \aCOunits$, while starburst-like galaxies have a much lower $\alpha_{\rm CO}\approx 0.8 \aCOunits$, where the CO measurement here corresponds to the lowest $J=1-0$ rotational transition. One traditional approach to generating an H$_2$ mass function is to measure CO(1-0), classify the galaxy into one of these categories, and use an appropriate factor typically assumed to be a constant among all galaxies in a class. Clearly this is quite simplistic, and it is more likely that there is a continuum of $\alphaCO$ values.

\citet{narayanan12a} presented a theoretical investigation into how $\alphaCO$ varies with galaxy properties. They used a suite of very high resolution isolated disk galaxy and merger simulations, together with CO line radiative transfer modeling, to directly connect the H$_2$ measured in their galaxies with the emergent CO(1-0) luminosity.  They find that the relation between $\alphaCO$ and galaxy properties is on average reasonably well described by:
\begin{equation}\label{eqn:XCO_narayanan}
    \alpha_{\rm CO} = \frac{20.6}{Z_{\rm H_2}\Sigma_{\rm H_2}^{0.5}}\;\; \aCOunits,
\end{equation}
where $Z_{\rm H_2}$ is the mass-weighted metallicity of the molecular gas in solar units, and $\Sigma_{\rm H_2}$ is the molecular mass surface density in ${\rm M}_\odot$pc$^{-2}$.

The quantities $Z_{\rm H_2}$ and $\Sigma_{\rm H_2}$ are calculable for each galaxy in our various simulations.  To compute $Z_{\rm H_2}$, we determine the H$_2$ fraction-weighted metallicity of all gas particles with a radius containing half the molecular gas ($R_{\rm H_2}$).  For  $\Sigma_{\rm H_2}$, we compute the projected H$_2$ surface densities from gas particles within $R_{\rm H_2}$ in the $z$ direction, although we checked in \simba\ that using the average of the $x$, $y$, and $z$ projections gives similar results.  Given these quantities, we use equation~\ref{eqn:XCO_narayanan} to compute $\alpha_{\rm CO}$ for each galaxy. This enables us to predict CO(1-0) luminosity function and scaling relations for direct comparison to the CO(1-0) luminosities in xCOLDGASS and other CO surveys.  We note that \citet{narayanan12a} recommends applying their formula locally within the ISM, but given the $\sim$kpc scale spatial resolution of our simulations this is often impractical.  Hence we compute these quantities within $R_{\rm H_2}$ to give a single $\alpha_{\rm CO}$ value for each galaxy.

\begin{figure}
	\includegraphics[width=0.48\textwidth]{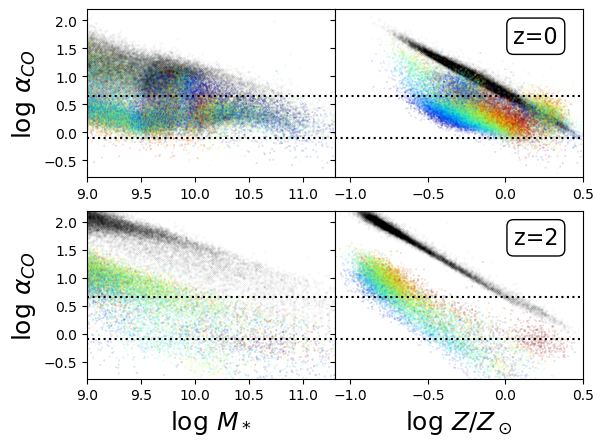}
	\vskip-0.1in
    \caption{$\alphaCO$ versus stellar mass (left panels) and metallicity (right), at $z=0$ (top panels) and $z=2$ (bottom), in \simba, using the \citet{narayanan12a} prescription. Individual galaxies are colour-coded by their location relative to the main sequence $\Delta_{MS}=-0.7\to +0.7$ (red to blue).  Black hexbins show the values using instead the \citet{Accurso:2017} prescription.  Horizontal dotted lines demarcate values usually assumed for starbursts ($\alphaCO=0.8$) and Milky Way-like galaxies ($\alphaCO=4.5$), for reference.}
    \label{fig:alphaCO}
\end{figure}

\begin{figure}
	\includegraphics[width=0.45\textwidth]{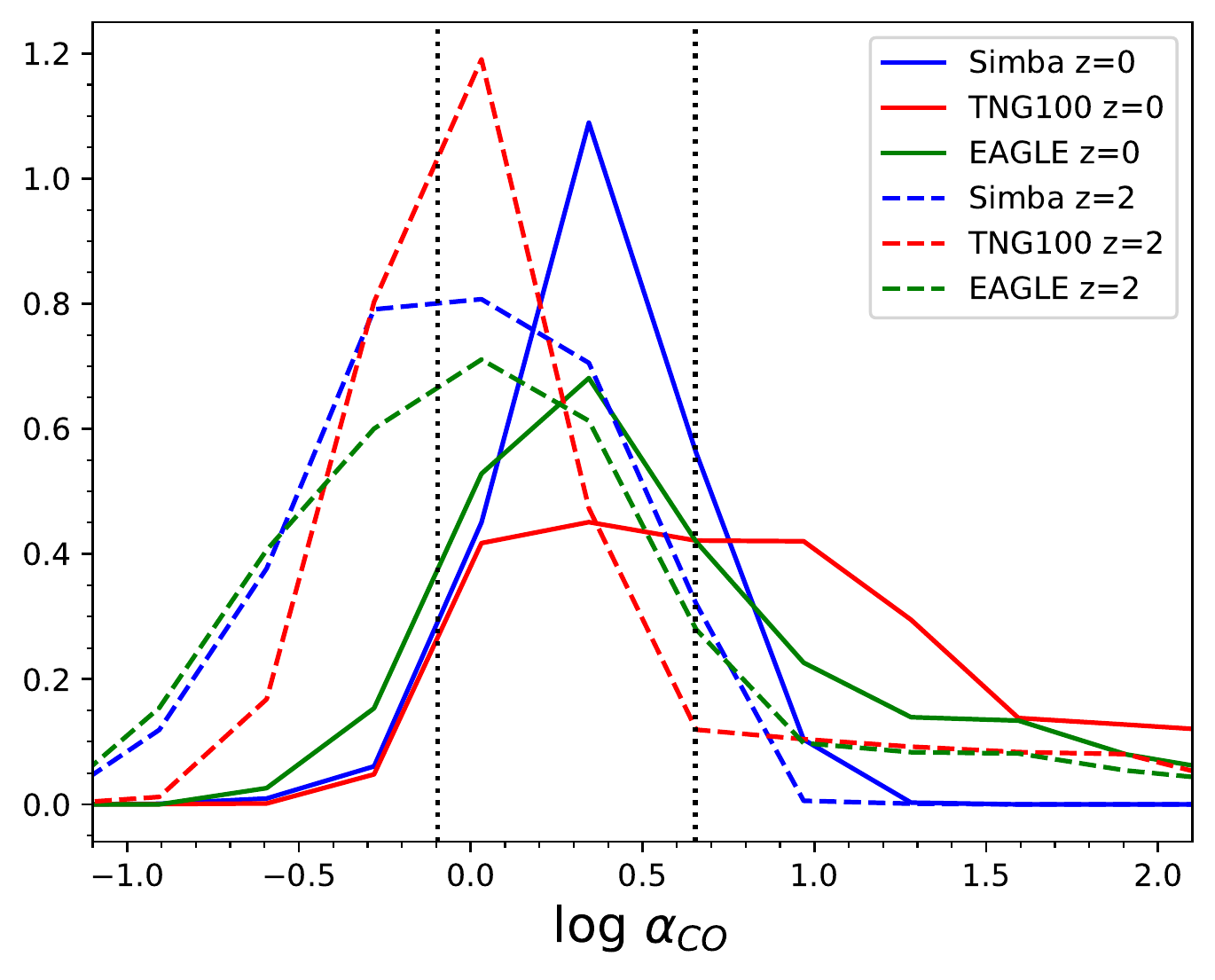}
	\vskip-0.1in
    \caption{Histograms of $\alphaCO$ for $M_*>10^{10}{\rm M}_\odot$ galaxies in \simba\ (blue), \TNG\ (red), and EAGLE (green), at $z=0$ (solid curves) and $z=2$ (dashed).  Vertical lines show values for starbursts ($\alphaCO=0.8$) and Milky Way-like galaxies ($\alphaCO=4.5$), for reference}
    \label{fig:alpha_comp}
\end{figure}

Fig.~\ref{fig:alphaCO} shows a scatter plot of $\alphaCO$ versus stellar mass (left panels) and metallicity (right panels), at $z=0$ (top) and $z=2$ (bottom), in \simba, computed using the \citet{narayanan12a} prescription (eq.~\ref{eqn:XCO_narayanan}).  Individual galaxies are colour-coded by their location relative to the star-forming main sequence, computed as a running median of SFR vs. $M_*$, ranging from reddest points having $\Delta_{MS}\leq -0.7$ to the bluest points with $\Delta_{MS}\geq 0.7$.  The black hexbin shading in the background shows the values computed instead using the \citet{Accurso:2017} method (eq.~\ref{eq:XCO_accurso}).  The horizontal dashed lines show reference values typically assumed for starbursts ($\alphaCO=0.8$) and Milky Way-like galaxies ($\alphaCO=4.5$).  Only galaxies with $M_{\rm H_2}>10^8{\rm M}_\odot$ (as well as our adopted stellar mass resolution limit of $M_*>10^9{\rm M}_\odot$) are shown, since otherwise they have too little molecular gas to reliably determine the quantities required to estimate $\alphaCO$.

Generally, $\alphaCO$ is anti-correlated with both stellar mass and metallicity.  The trend with metallicity is stronger, reflecting the inverse metallicity dependence in equation~\ref{eqn:XCO_narayanan}.  At a given metallicity, low-sSFR galaxies have higher values of $\alphaCO$, since they are gas-poor with lower H$_2$ surface densities. The trend with mass is shallower than that with metallicity at $z=0$, because lower-mass galaxies have lower metallicity, but this is partly counteracted by their higher gas surface densities.  It is notable that \simba\ seems to under-predict the value of $\alphaCO$ in a Milky Way-like star-forming galaxy relative to the nominal value of $\alphaCO=4.5\aCOunits$; this is also true for EAGLE.  This would result in an over-prediction of CO luminosities for a given $M_{\rm H_2}$; \romeel{we quantify the implications of this for the CO luminosity function below.}

At $z=2$, the overall values \romeel{using the \citet{narayanan12a} prescription}\ are lower than at $z=0$ for a given mass or metallicity.  This is because galaxies are more compact and gas-rich at high redshifts~\citep[e.g.][]{Appleby:2019}, leading to higher $\Sigma_{\rm H_2}$; \romeel{while the  metallicities are also lower~\citep{Dave:2019}, the relatively weak dependece on $Z_{H2}$ is more than compensated by the increased surface density}.  

The black hexbins in the background show $\alphaCO$ values computed using equation~\ref{eq:XCO_accurso} from \citet{Accurso:2017}, for comparison.  At $z=0$ for massive ($M_*\ga 10^{10}{\rm M}_\odot$) galaxies, the values from the two methods are similar, \romeel{though with a slight trend towards higher $\alphaCO$ using this method.} At lower masses, the \citet{Accurso:2017} prescription yields more significantly higher values, since it is more strongly dependent on metallicity, which is also reflected in its tighter relation with metallicity in the right panel. At high redshifts, equation~\ref{eq:XCO_accurso} gives implausibly high values, owing to the significantly lower metallicities that is not mitigated by the higher gas surface densities as in the \citet{narayanan12a} prescription. 
This is perhaps not surprising, given that the \citet{Accurso:2017} prescription is based on $z\approx 0$ observations.  For the remainder of this work, we will use the \citet{narayanan12a} prescription, as it appears to yield a more plausible redshift evolution owing to accounting for both structural and metallicity changes.

To compare the $\alphaCO$ values among the different simulations, we show in Fig.~\ref{fig:alpha_comp}
histograms of $\alphaCO$ at $z=0,2$ (solid and dashed lines, respectively) for galaxies with $M_*>10^{10}{\rm M}_\odot$ from \simba\ (blue), \TNG\ (red), and EAGLE (green).  Reference lines for typical starburst and MW values are indicated by the vertical dotted lines.

\simba\ and EAGLE both predict a median value of $\alphaCO=2.9 \aCOunits$ at $z=0$, dropping to $\approx 1$ at $z=2$.  A typical dispersion of $\alphaCO$ is $\approx 0.5$~dex, which is consistent with the spread seen in Fig.~\ref{fig:alphaCO} for \simba. This shows that the assumption of a constant $\alphaCO$ value even among relatively massive star-forming galaxies may be a poor one.  Meanwhile, \TNG\ shows generally higher values of $\alphaCO$, with a median value of $\alphaCO\approx 6$ at $z=0$ with rapid evolution to $\alphaCO\approx 1.5$ at $z=2$, and an even larger dispersion.

It is worth pointing out that the $\alphaCO$ prescription developed by \citet{narayanan12a} used simulations with star formation and feedback prescriptions that are different to any of the simulations considered here, and also were not cosmologically situated.  Since $\alphaCO$ likely depends on the structure and distribution of molecular clouds within the ISM, it could be sensitive to such choices.  For instance, the same procedure of running very high resolution zoom versions using our three simulations' own star formation and feedback prescriptions and applying a CO radiative transfer code could yield substantially different fitting formulae for $\alphaCO$ in each case. While it is beyond the scope to investigate this here, the variations in $\alphaCO$ among the different simulations even when using the same underlying fitting formula highlight the importance of being able to predict this quantity more accurately in $\sim$kpc scale cosmological simulations if one wants to more robustly compare such simulations to CO observations.

Overall, our computed values of $\alphaCO$ broadly follow expected trends of being around the Milky Way value in massive star-forming galaxies today, shifting towards more starburst-like values at high redshifts. There is, however, no bimodality in the $\alphaCO$ distribution, indicating that using a bimodal $\alphaCO$ value based on galaxy classification may be too simplistic. Moreover, the large spread in $\alphaCO$ at a given $M_*$ or metallicity suggests that using a single value, regardless of what it is, may be a dangerous assumption.  This is particularly true when examining counting statistics such as a mass function, where the scatter in $\alphaCO$ could scatter more numerous low-$M_{\rm H_2}$ galaxies up to high $L_{\rm CO(1-0)}$ values, thereby increasing $M^\star_{\rm H_2}$ over what one would infer from assuming a constant $\alphaCO$.  In order to examine such effects more quantitatively, we next compare the resulting CO(1-0) luminosity function with $\alphaCO$ computed as above among our various simulations, and compare these to observations from $z\sim 0-2$.

\subsection{The CO Luminosity Function}\label{sec:COLF}

With a prescription for computing $\alphaCO$ in hand, albeit with its substantial attendant uncertainties, we can now move the comparison of molecular gas into the observational plane.  It is particularly interesting to relate the comparative trends seen for the COLF to the analogous trends seen for the H2MF from the previous section -- if $\alphaCO$ was a robust and well-determined quantity, then the general trends between these should mirror each other, but we will see that there are significant differences. Moreover, we can also engage in more direct comparisons to observations out to higher redshifts.  Thanks to recent surveys such as ASPECS and COLDz, the CO(1-0) luminosity function (COLF) has now been measured out to $z\ga 2$, to go along with the improved recent low-redshift determination from xCOLD~GASS~\citep{Saintonge:2017}.  In this section we compare our simulated COLFs to each other and to these observations, to assess how well current models do at reproducing data and understand how robust these comparisons are.

\begin{figure}
	\includegraphics[width=0.48\textwidth]{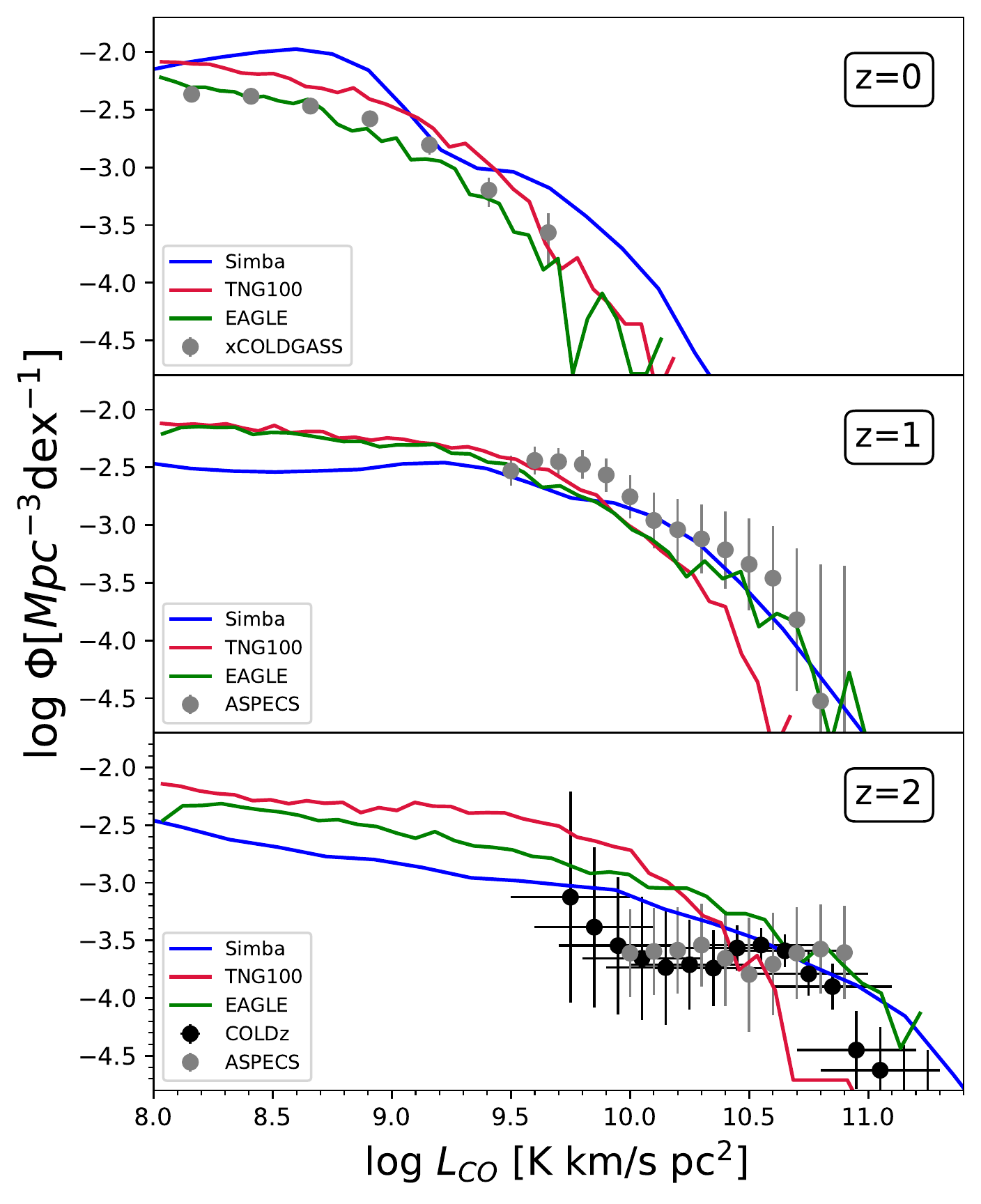}
	\vskip-0.1in
    \caption{CO(1-0) luminosity functions for \simba\ (blue), \TNG\ (red), and EAGLE (green) at $z=0,1,2$ (top to bottom).  These are computed using \citet{narayanan12a} prescription for $\alphaCO$; the results from using a constant $\alphaCO$ at the mean value instead results in the dotted lines.  Observations from xCOLD~GASS at $z=0$ \citep{Saintonge:2017}, ASPECS at $z\approx 1,2$~\citep{Aravena:2019}, and COLDz at $z\approx 2$~\citep{Riechers:2019} are shown in black and grey.  \TNG\ and EAGLE well reproduce the COLF despite widely different H2MF, while \simba\ overproduces the COLF like the H2MF.  \simba\ and EAGLE are able to reproduce observations of high-$L_{\rm CO1-0}$ galaxies at $z\ga 1$.}
    \label{fig:COLF}
\end{figure}

Fig.~\ref{fig:COLF} shows the CO(1-0) luminosity function at $z=0,1,2$ (top to bottom) for \simba\ (blue), \TNG\ (red), EAGLE (green).  Also shown are various observational determinations: The grey points at $z=0$ are the observations from xCOLD~GASS~\citep{Saintonge:2017}, while at $z\approx 1,2$ we show observations from ASPECS~\citep{Aravena:2019} and COLDz~\citep{Pavesi:2018,Riechers:2019}.   

The $z=0$ COLF, with $\alphaCO$ computed individually for each simulated galaxy, gives a qualitatively different picture in comparison to observations.  Firstly, all the models are now significantly closer to the observations. For instance, EAGLE has gone from being extremely deficient at high $M_{\rm H_2}$, to agreeing very well for $L_{\rm CO1-0}$. 
\TNG\ showed a milk overproduction in the H2MF at all masses, but now agrees quite well with the COLF, thanks to its typically higher values of $\alphaCO$.
This illustrates that the assumptions about $\alphaCO$ qualitatively impact simulations constraints based on the H2MF. 

At higher redshifts, the qualitative evolution among the models mimics that seen for the gas mass functions:  EAGLE and \simba\ have strong positive luminosity evolution out to high redshifts, while \TNG's evolution is also positive (owing to its lower $\alphaCO$ values at high-$z$) but much weaker than in the other two simulations. The net result is that \TNG\ has a difficult time reproducing the very high $L_{\rm CO1-0}$ values seen in galaxies at $z\sim 1-2$, and tends to over-produce low-$L_{\rm CO1-0}$ systems.  A similar failing of galaxy formation models was noticed from a comparison to semi-analytic models done in \citet{Riechers:2019}, \romeel{and similarly \citet{Popping:2019} found that \TNG\ under-predicted the high-luminosity end at $z>1$}.  As is often the case, the claimed discrepancy is quite model dependent.  \simba\ and EAGLE are well able to produce $L_{\rm CO1-0}\sim 10^{11}$~K~$\kms$~pc$^2$ systems at $z\sim 2$, as observed.  The key is their low $\alphaCO$ values, typically $\alphaCO\sim 1\aCOunits$.
We note that \simba\ reproduces the mass-metallicity relation at $z\sim 2$~\citep[as well as $z\sim 0$;][]{Dave:2019}, so the low $\alphaCO$ values are not due to over-enriched galaxies.

\begin{figure}
	\includegraphics[width=0.48\textwidth]{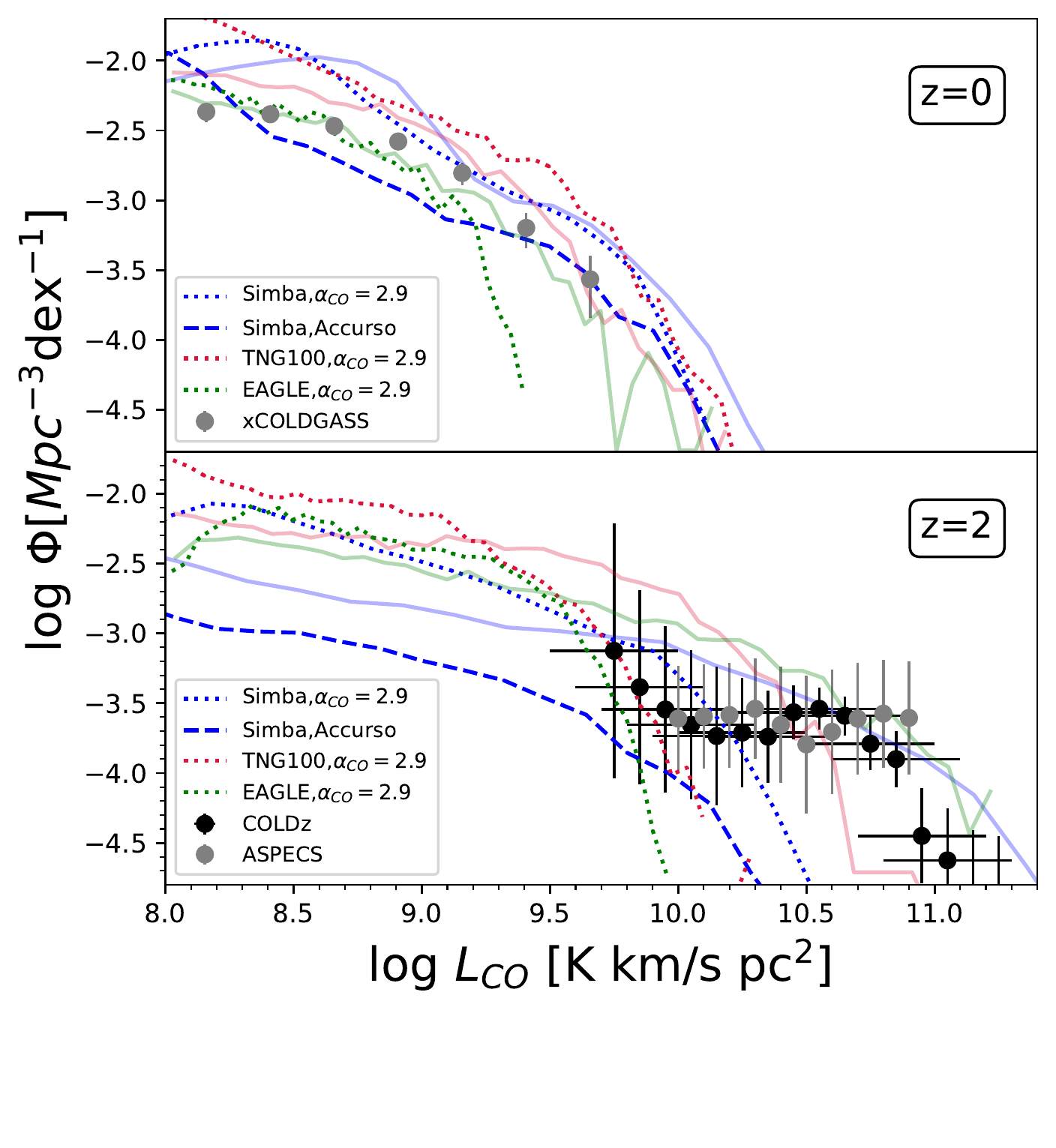}
	\vskip-0.1in
    \caption{CO(1-0) luminosity functions for \simba\ (blue), \TNG\ (red), and EAGLE (green) at $z=0,2$ (top to bottom).  \romeelnew{The different line types show the results when computed using various $\alphaCO$ prescriptions. The faint solid lines show the results using the \citet{narayanan12a} prescription,  reproduced from Figure~\ref{fig:COLF}.  The dotted lines for each simulation show the results assuming a constant $\alphaCO=2.9\; \aCOunits$, which is a typical average value at $z=0$.  Finally, the dashed blue line shows the COLF for \simba\ assuming the \citet{Accurso:2017} $\alphaCO$ prescription, which produces unphysically high values of $\alphaCO$ at $z=2$.  For reference, observations are shown as in Figure~\ref{fig:COLF}.  The assumption of a constant $\alphaCO$ makes an especially large difference for the bright end at $z=2$.}}
    \label{fig:COLF_comp}
\end{figure}

\romeel{
To illustrate the sensitivity of these predictions to assumptions about $\alphaCO$, Figure~\ref{fig:COLF_comp} shows a comparison of these models with different assumptions about $\alphaCO$, at $z=0$ (top) and $z=2$ (bottom).  For \simba, we show with blue dotted and dashed curves the results of using a constant $\alphaCO= 2.9~\aCOunits$ and the \citet{Accurso:2017} $\alphaCO$ prescription, respectively.  For \TNG\ and EAGLE shown in red and green, the dotted lines show a constant $\alphaCO= 2.9~\aCOunits$.  For comparison, the semi-transparent solid lines reproduces the results from Figure~\ref{fig:COLF}, and the observations shown there are also reproduced.

For \simba\ at $z=0$, using the median $\alphaCO$ (dotted line) rather than the full spread in values does not yield a much different COLF.  At $z=2$, however, there is a large difference, as using a constant $\alphaCO$ strongly underpredicts the bright end.  This illustrates the importance of including a distribution of $\alphaCO$ values for comparing to observations.  Meanwhile, the \citet{Accurso:2017} prescription yields a somewhat lower COLF at $z=0$, owing to its generally higher $\alphaCO$ values.  At $z=2$, the likely unphysically high $\alphaCO$ values predicted in this prescription results in a much poorer agreement with data.

For EAGLE and \TNG, the story is similar:  At $z=0$, using a constant $\alphaCO$ results in mildly lower COLFs, but at $z=2$, the difference is very pronounced, and as with \simba\ tends to strongly truncate the bright end of the COLF.  The assumption of a constant $\alphaCO$ may thus be a major reason why \citet{Riechers:2019} and \citet{Popping:2019} found that models could not reproduce the bright end of the high-redshift COLF.
}

Clearly, independent constraints on $\alphaCO$ at both low and high redshifts would be highly valuable in order to conduct a robust comparison between the observed and simulated COLFs. This could come from direct observations~\citep[e.g.][]{Accurso:2017}, or else from sophisticated higher-resolution simulations including CO line radiative transfer, such as with S\'IGAME~\citep{Olsen:2017}.  It is possible that the $\alphaCO$ values coming from the \citet{narayanan12a} prescription are systematically discrepant in one or more of our simulations, which could then either indicate a failing of that model, or the inapplicability of the \citet{narayanan12a} prescription for that model. There is thus substantial effort still needed in order to be able to utilise the COLF as a robust constraints on galaxy formation models.

Overall, it is encouraging that all our models better reproduce the $z\approx 0$ COLF than the H2MF, since the former is the more direct observable.  At higher redshifts, at least some current galaxy formation models have no difficulty forming galaxies with high $L_{\rm CO1-0}$ values at $z\sim 1-2$ -- the evolution predicted in EAGLE is in very good agreement with observations, \simba's evolution is still quite reasonable compared to data, while \TNG\ has significant difficulties generating high-$L_{\rm CO1-0}$ systems at high redshifts despite its rapid downwards evolution of $\alphaCO$.  EAGLE's good agreement with the COLF was also noted in \citet{Lagos:2015}, despite using a different metallicity-dependent $\alphaCO$ prescription.  Nonetheless, all these conclusions are highly sensitive to assumptions regarding $\alphaCO$.  To make progress, this must be independently constrained either observationally and/or theoretically in order to properly assess whether galaxy formation models match observations of molecular gas in galaxies across cosmic time.

\subsection{Gas Fraction Scaling Relations}

\begin{figure*}
	\includegraphics[width=1.0\textwidth]{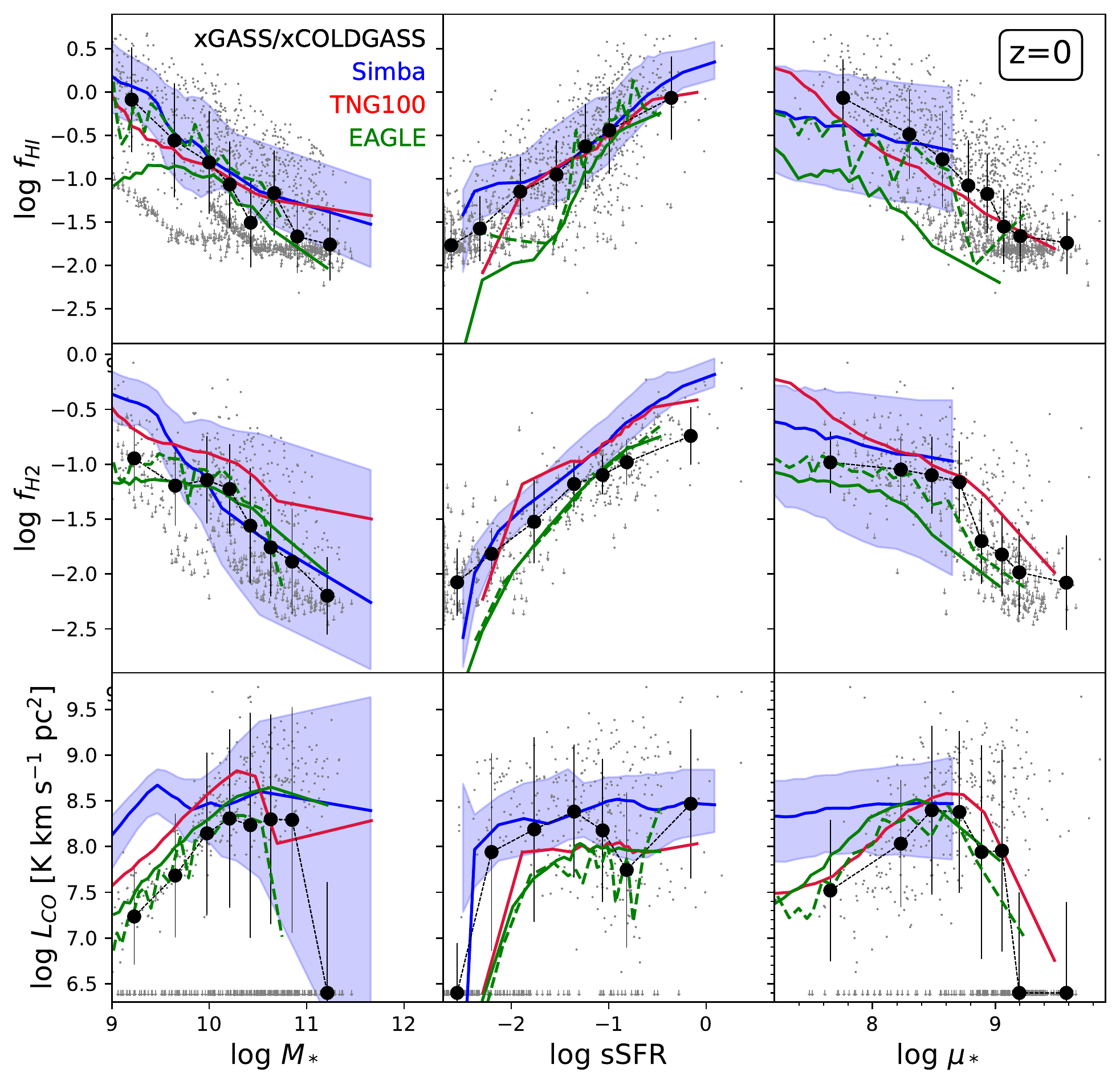}
	\vskip-0.1in
    \caption{Gas fraction scaling relations at $z=0$ in \simba\ (blue), \TNG\ (red), EAGLE (green solid), and EAGLE-Recal (green dashed). Top row shows atomic gas fraction $M_{\rm HI}/M_*$, middle shows molecular gas fraction $M_{\rm H_2}/M_*$, and bottom row shows $L_{\rm CO1-0}$ with $\alphaCO$ computed using the \citet{narayanan12a} prescription. These are shown as a function of stellar mass (left panels), specific SFR (middle), and stellar surface density (right).  Lines shown the running median for each scaling relation; in each case, the bins have been chosen to hold approximately an equal number of galaxies from within each dataset. Blue shaded region shows the 16--84\% spread around the median in \simba.
    Grey points show results from xGASS for \HI\ data and xCOLD~GASS for H$_2$ and CO data (non-detections shown as downward arrows at their upper limits), with the running median shown as the black points with errorbars indicating the $1\sigma$ spread about the median \romeel{including non-detection}.  All simulations broadly reproduce the trends in gas fractions, but significant discrepancies are seen for \simba\ in molecular gas at low $M_*$ and all sSFR, for \TNG\ in molecular gas at high $M_*$, and for EAGLE in \HI\ at low masses which is much improved in EAGLE-Recal.}
    \label{fig:gass}
\end{figure*}

We have seen that our three galaxy formation simulations qualitatively reproduce the distribution functions of cold gas and their measures in galaxies, but there are also significant discrepancies in each case. To investigate the successes and failures in more detail and isolate the galaxy population(s) responsible, it is instructive to examine scaling relations of gas content versus global galaxy properties, which is what we do here.

Fig.~\ref{fig:gass} shows a montage of scaling relations for our simulations, compared to observations.  The $y-$axis panels shows the quantities $\fneut=\mhi/M_*$, $\fmol=M_{\rm H_2}/M_*$, and $L_{\rm CO(1-0)}$ as computed assuming the \citet{narayanan12a} prescription for $\alphaCO$.  The $x-$axis quantities are the stellar mass $M_*$, the specific star formation rate (sSFR), and stellar mass surface density (computed within the half stellar mass radius) $\mu_*$.  In each panel, we show a running median for \simba\ (blue), \TNG\ (red), EAGLE (green solid), and EAGLE-Recal (green dashed), with the spread from the 16th to 84th percentile shown as the shaded blue region for \simba.  The underlying grey points show the observations from xGASS (for \HI) and xCOLD~GASS (for H$_2$ and CO) with downwards arrows indicating upper limits.  A running median is shown as the black points with the errorbars indicating the $1\sigma$ spread around the median. \romeel{The medians are taken over all data points including non-detections or gas-free galaxies; using the median rather than the mean avoids any ambiguity regarding the values for the upper limits in the observations.  The results are not significantly different if we compute the running mean instead}.  The bins are chosen to have roughly equal numbers of galaxies in each.

The leftmost column shows the relations versus $M_*$. In the simulations, stellar mass is generally the most accurately predicted quantity, hence $M_*$ scaling relations are likely the most robust trends predicted by the models.  All models predict falling $\fneut$ and $\fmol$ with increasing $M_*$, and a slow rise in $L_{\rm CO1-0}$ with $M_*$ in the star-forming regime and a quick drop in the most massive (generally quenched) galaxies.  The variance around the median in \simba\ is about 0.4 dex in \HI, and increases towards high $M_*$ for the molecular and CO trends.  These broadly mimics the corresponding observational trends, but there are notable discrepancies.

For $\fneut$, \simba\ and \TNG\ agree reasonably well at most masses, reflecting their good agreement with the HIMF as seen in Fig.~\ref{fig:HIMF}, but somewhat overpredict the atomic fractions at the highest masses.  
EAGLE, meanwhile, is an interesting case -- it strongly underpredicts the HIMF, yet the $\fneut$ is only mildly low, except for a stronger drop at the lowest $M_*$.  This occurs because EAGLE has a significantly larger fraction of galaxies across all masses with little or no \HI, which impacts the counts more than the median values.  EAGLE-Recal, meanwhile, has higher $\fneut$ at all masses, but in particular follows the observations  more closely at $M_*\la 10^{10}{\rm M}_\odot$, the combination of which produces an HIMF for EAGLE-Recal that is in good agreement with observations. As discussed in Section \ref{sec:HIMF}, we speculate that the significant difference between EAGLE and EAGLE-Recal is a consequence of the stronger resolution dependence of EAGLE's subgrid feedback model relative to those used by \simba\ and \TNG.

$\fmol$ similarly shows an overall falling trend in all models. \TNG\ overproduces the molecular fractions particularly in low- and high-$M_*$ galaxies, and likewise shows an upwards deviation in $\fmol$ towards high $M_*$ like that seen for $\fneut$, indicating that this represents a true bump in the overall cold gas in massive systems as opposed to some artifact of \HI--H$_2$ separation.
\romeel{The overprediction, particularly at the massive end, is subject to uncertainties regarding apertures; \citet{Diemer:2019} and \citet{Popping:2019} obtained significantly better agreement owing to their use of a smaller aperture.}

\simba, meanwhile, overproduces $\fmol$ particularly at low masses, suggesting that the excess seen in the H2MF comes from dwarfs that are overly molecular gas-rich.  \romeel{Given \simba's large aperture, it is subject to similar aperture caveats as \TNG.  Curiously, despite matching the H2MF and the GSMF fairly well, \simba\ systematically overproduces the H2 fractions.}

EAGLE shows the best agreement in the slope of $\fmol(M_*)$, although it is slightly low.  Hence for the galaxies that have molecular gas, EAGLE does a good job of reproducing their gas fractions. EAGLE-Recal is very similar to EAGLE, showing that molecular fractions are less resolution-sensitive than atomic fractions in EAGLE.  

Both \TNG\ and \simba\ produce at least some quite massive galaxies with significant H$_2$ even though those galaxies are generally quenched, which is also seen in some observed systems~\citep[e.g.][]{Davis:2019}.  It remains to be seen if there is statistical agreement with observations since xCOLD~GASS is not a sufficiently large sample to include such rare objects, and current observations of molecular gas in massive galaxies are limited to heterogeneously selected samples.  In \simba, such gas typically has very low star formation efficiency and lies below the Kennicutt-Schmidt relation~\citep{Appleby:2019}, likely owing to its diffuse distribution.  

\romeel{We can further compare these to the \mufasa\ gas scaling predictions shown in Figure~5 of \citet{Dave:2017a}.  The $\fmol$ relation is fairly similar, since the H$_2$ formation model has remained the same, but \simba\ produces more molecular gas in low-mass galaxies.  The $\fneut$ predictions are also fairly similar, but \simba\ produces somewhat more \HI\ at the highest masses.  Likely this occurs because \mufasa's quenching feedback mechanism explicitly heated all the ambient gas in high-mass halos.  \mufasa\ thus over-suppressed \HI\ in massive galaxies, while \simba\ slightly under-suppresses it.}

Looking at $L_{\rm CO1-0}$, we see that the impact of the variations in $\alphaCO$ result in somewhat different trends relative to that for $\fmol$.  As in the COLF, \simba\ clearly overproduces $L_{\rm CO1-0}$, while EAGLE and \TNG\ do generally better, the latter owing to its significantly higher values of $\alphaCO$.  \simba\ and EAGLE show a sharp decline in CO luminosity at $M_*\ga 10^{11}{\rm M}_\odot$, similar to that seen in xCOLD~GASS, while \TNG\ continues to show typically high $L_{\rm CO1-0}$ out to large stellar masses.  In \simba, despite there being some molecular gas in these galaxies, the H$_2$ surface densities are generally low, which increases $\alphaCO$ and thus decreases the CO luminosity.  Nonetheless, the very most massive galaxy in \simba\ is relatively CO-bright.

The second column depicts the trends versus sSFR.  Again, all simulations qualitatively produce the observed trends of increasing cold gas fractions with sSFR, but only a weak trend with $L_{\rm CO1-0}$. The trends in individual simulations mirror that seen for $M_*$: \TNG\ and \simba\ match $\fneut$ and are somewhat high in $\fmol$, reflecting the trends seen in the mass functions.  For EAGLE, the origins of the discrepancies in the mass functions become clearer in this plot:  EAGLE produces too little neutral gas in low-sSFR systems, and generally shows a steeper slope of either \HI\ or H$_2$ gas fraction versus sSFR versus observations, indicating that EAGLE's model over-suppresses cold gas in green valley galaxies.  EAGLE-Recal simply lacks many galaxies in this green valley regime, but for the few that are there, it seems to produce significantly higher $\fneut$.  Meanwhile, the $L_{\rm CO1-0}$ trends represents the competing effects of lower sSFR objects generally having lower molecular fractions, but also being larger systems overall.  All models achieve this balance naturally, although as before the amplitudes vary somewhat.  All models also produce a sharp drop in $L_{\rm CO1-0}$ for the quenched systems as observed, more dramatic than that for $\fmol$. In \simba\ owes to a higher $\alphaCO$ in such systems from the lower molecular gas surface densities, and this appears to occur naturally in the other simulations as well.

The rightmost column shows trends versus the stellar surface density $\mu_*$.  Both data and models show dropping cold fractions with increasing $\mu_*$, as higher $\mu_*$ galaxies tend to have lower sSFR and therefore less cold gas relative to stars.  The slope of the trend in all models generally reproduces observations.   However, the amplitudes vary substantially. \TNG\ produces good agreement with all properties as a function of $\mu_*$, while EAGLE tends to under-produce the gas contents,
and \simba\ predicts reasonable gas fractions but generally has a flatter trend with $\mu_*$ than observations.  \romeel{The maximum $\mu_*$ reached in each model is substantially different, which is directly tied to }~numerical resolution: \TNG\ has the highest resolution, then EAGLE, then \simba, which has $\sim 15\times$ lower mass resolution than \TNG\ owing to its larger volume and lower number of gas elements. This is also seen by comparing EAGLE to EAGLE-Recal; the latter produces higher $\fneut$ values and also higher $\fmol$, even though the molecular fractions are a given $M_*$ or sSFR are not resolution-dependent.  One interpretation is thus that \TNG's resolution is necessary in order to properly resolve the inner structure of the stellar component.  If so, this means predicting inner structural quantities as stellar mass surface density, bulge fractions, Sersic indices, etc., is quite computationally demanding and subject to a careful assessment of resolution convergence effects.  However, we will show in \S\ref{sec:simba_agn} that \simba\ with AGN feedback off produces a plethora of very high $\mu_*$ galaxies, even at the same resolution.  
Hence it is certainly possible to produce galaxies with high $\mu_*$ at \simba's resolution, which suggests that instead it may be \simba's AGN feedback scheme that prevents galaxies from having sufficiently high stellar densities.

Overall, all our simulations broadly reproduce trends versus stellar mass, surface density, and specific SFR, but there are notable variations and discrepancies that highlight specific differences reflected in the earlier mass and luminosity functions.  \TNG\ and \simba\ tend to do slightly better on \HI\ fractions, while EAGLE does somewhat better on molecular gas fractions.  Stellar surface densities provide a novel constraint on feedback regarding the spatial distribution of stellar growth in simulated galaxies, albeit such measures may be more sensitive to numerical effects.  As observations improve both at $z=0$ and in the distant universe, it is clear that gas scaling relations will provide complementary constraints and insights into the robustness and validity of the input physics in cosmological galaxy formation simulations.

\subsection{\simba\ AGN feedback variants}\label{sec:simba_agn}

\begin{figure*}
	\includegraphics[width=0.9\textwidth]{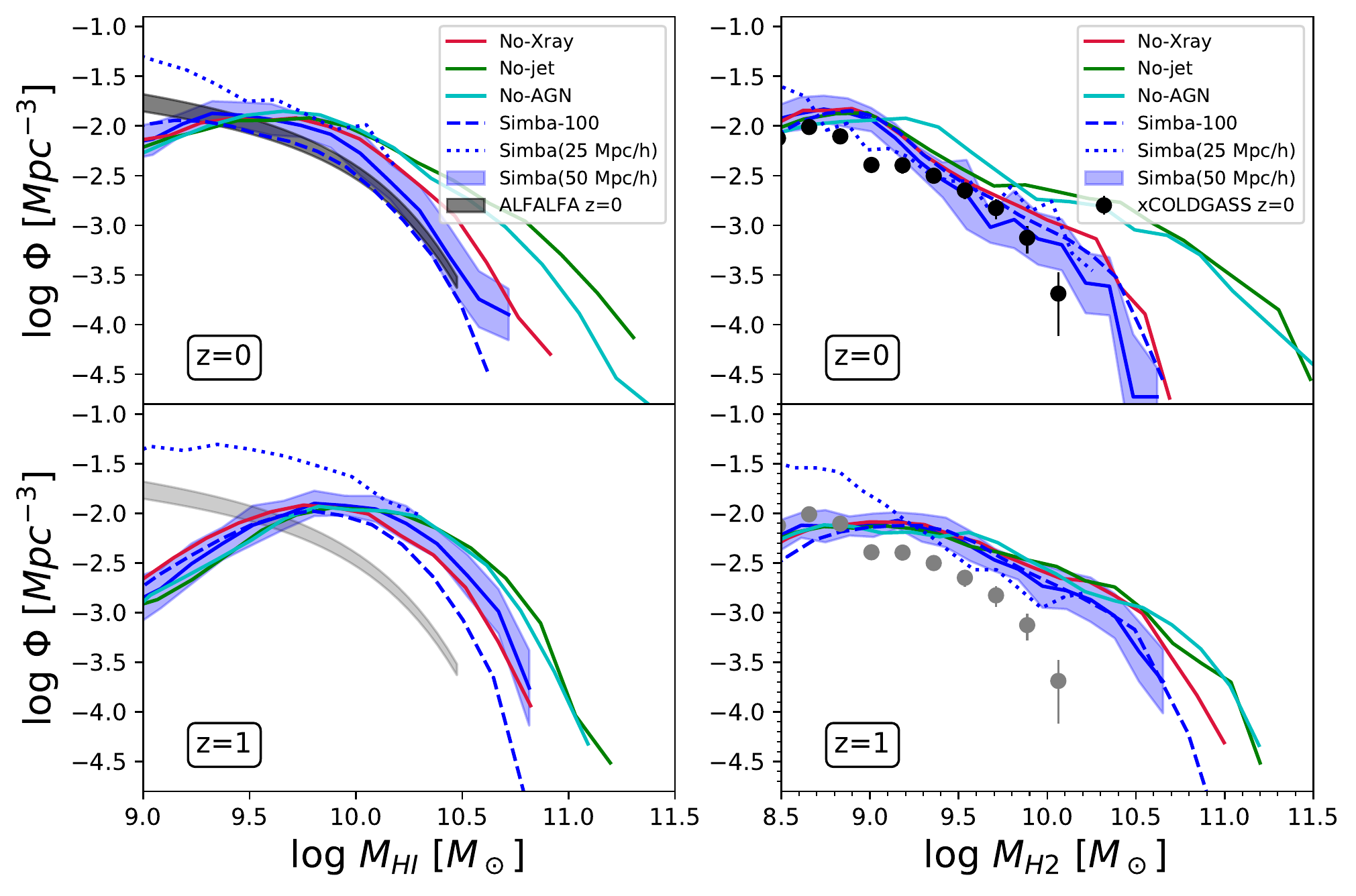}
	\vskip-0.1in
    \caption{HIMF (left) and H2MF (right) at $z=0,1$ (top, bottom) for the four AGN feedback variants: full \simba\ (blue), no-Xray (red), no-jet (green), and no-AGN (cyan).  These are run in $50\hmpc$ volumes; the corresponding \simba\ $100\hmpc$ volume results are shown as the blue dashed line.  Observations from ALFALFA~\citep{Jones:2018} and xCOLD~GASS~\citep{Fletcher:2020} are shown for reference.  The dominant difference comes from the inclusion of AGN jet feedback, which strongly suppresses cold gas in massive galaxies at $z\la 1$.}
    \label{fig:mfgas}
\end{figure*}

\begin{figure*}
	\includegraphics[width=0.9\textwidth]{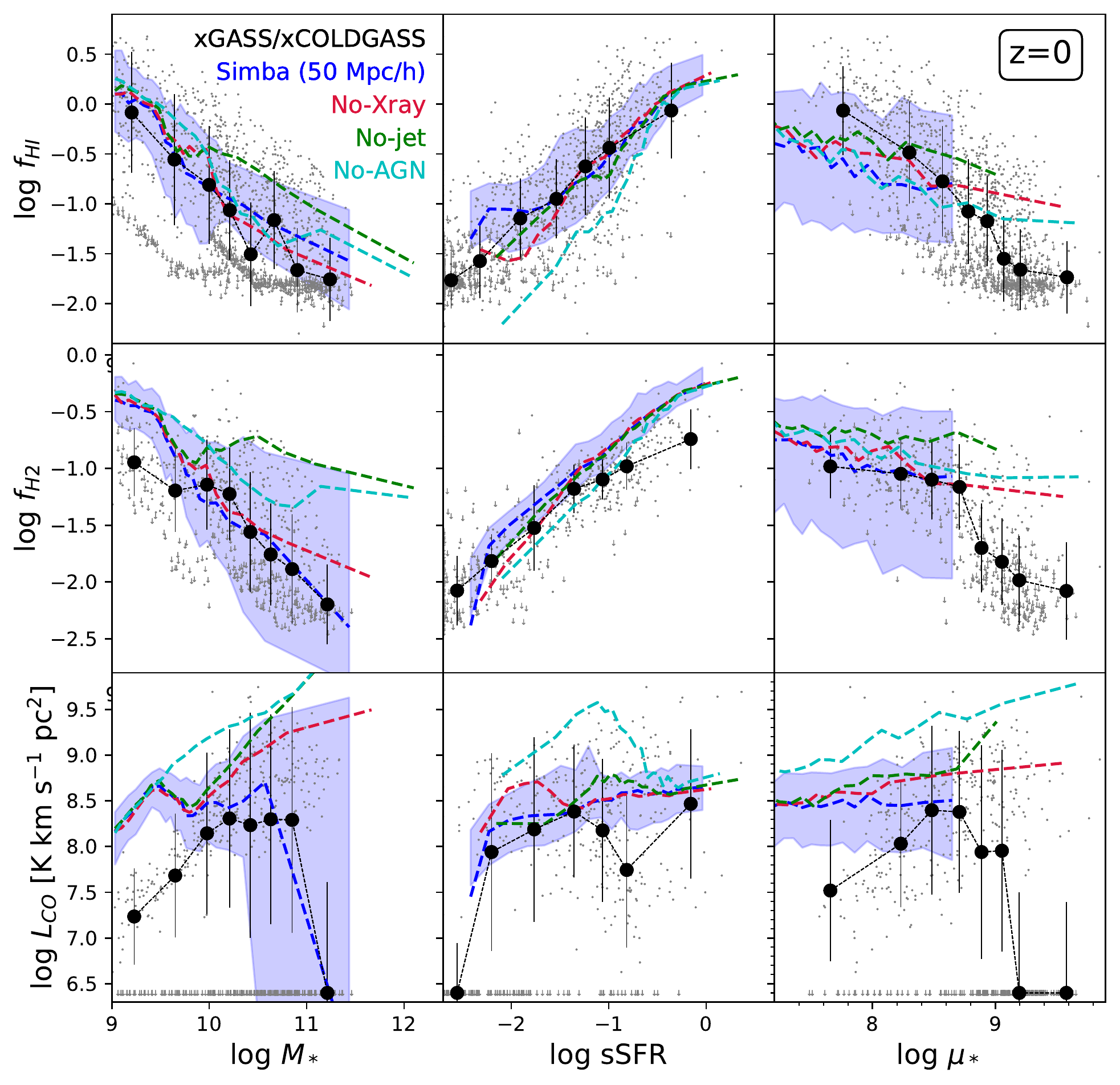}
	\vskip-0.1in
    \caption{Gas scaling relations at $z=0$ as in Fig.~\ref{fig:gass}, for $50\hmpc$, $2\times 512^3$ variants of \simba\ with different AGN feedback modules included.  The blue line shows the running median of these scaling relations for the full \simba\ model, red (`No-Xray') shows turning off the X-ray feedback only, green (`No-jet') shows turning off the jets and X-ray feedback, and cyan (`No-AGN') showing further turning off the radiative feedback.  xGASS and xCOLD~GASS data are also shown as in Fig.~\ref{fig:gass}.  AGN feedback has no impact on the low-$M_*$ or high-sSFR galaxies, but in particular, jet AGN feedback strongly lowers the gas content in massive galaxies to bring them into better agreement with observations.}
    \label{fig:gass_agn}
\end{figure*}

AGN feedback is a key ingredient in modern galaxy formation models in order to quench massive galaxies~\citep{Somerville:2015}.  In terms of the cold gas content, the predominant effect is to eject and/or prevent gas accretion, owing to energy injection from the AGN.  Thus it is expected that models predict low cold gas contents in massive galaxies, as we have seen above.  Yet AGN feedback can come in many different forms~\citep{Sturm:2011,Maiolino:2012,Heckman:2014}, such as radiatively-driven winds off the accretion disk, relativistic jets, and photo-heating of surrounding gas.  As such, it is interesting to know which type of AGN feedback is responsible for lowering the cold gas content.  In \simba, we have a particularly interesting model that employs three different types of AGN feedback, broadly following the three feedback modes described in \citet{Heckman:2014}.  Understanding how these different types of AGN feedback impacts galaxy cold gas content, particularly in massive galaxies, may provide useful insights that would help guide and constrain quenching models.

In this section, we examine the impact of AGN feedback on the cold gas content of galaxies in \simba.  As outlined in \S\ref{sec:simba}, the three different forms of AGN feedback in \simba\ are jet feedback at low $\fedd\la 0.02$, radiative feedback at high $\fedd$, and X-ray feedback at low $\fedd$ and gas fractions~\citep[see][for full details]{Dave:2019}.  The first two feedback modes are bipolar and purely kinetic, while the X-ray feedback is spherical and mostly kinetic, though typically of much lower strength.  The radiative feedback mode corresponds to what is often referred to as `quasar mode' feedback, and is designed to qualitatively model outflows of warm ionised~\citep{Perna:2017b} and/or cold molecular~\citep{Maiolino:2012} gas at speeds of many hundreds of km/s.  The jet mode aims to represent the impact of collimated relativistic radio jets, often called `radio mode' feedback, that eject hot plasma out to tens to hundreds of kpc before imparting their energy into the surrounding medium.  X-ray feedback is motivated more fully in \citet{Choi:2012} as high-energy radiation pressure that imparts momentum onto the gas surrounding the black hole; unlike the other modes, this mode is spherical, and typically generates a relatively modest outward push of a few hundred km/s in the gas closest to the black hole, with a strength that scales as the inverse square of the distance.

The \simba\ suite contains $50\hmpc$ box size, $2\times 512^3$ element variants where each of these AGN feedback mechanisms is turned off in turn, as follows:
\begin{itemize}
    \item ``Simba'' -- Full AGN feedback physics.
    \item ``No-Xray'' -- Only X-ray feedback turned off.
    \item ``No-jet'' -- X-ray and jet feedback turned off.
    \item ``No-AGN'' -- All AGN feedback turned off.
\end{itemize}
All simulations are run from the same initial conditions, and all other input physics is unchanged.  These should thus be regarded as numerical experiments to isolate the impact of each feedback mechanism, rather than realistic variations of galaxy formation models.  For instance, only the full \simba\ model accurately reproduces the observed stellar mass function~\citep{Dave:2019}.

Fig.~\ref{fig:mfgas} shows the \HI\ (left panels) and H$_2$ (right) mass functions, at $z=0$ (top panels) and $z=1$ (bottom) for the $50\hmpc$ \simba\ run (blue), No-Xray (green), No-jet (red), and No-AGN (cyan).  The blue shading shows the cosmic variance estimated over 8 simulation sub-octants in the full physics run; the other models show similar variance but are not shown for clarity.  The dotted blue line shows the results from a small-volume ($25\hmpc$) version of \simba\ with $8\times$ better mass resolution.  The dashed blue line shows the $100\hmpc$ \simba\ results reproduced from Figs~\ref{fig:HIMF} and \ref{fig:H2MF}.   Observations from ALFALFA (left) and xCOLD~GASS (right) are also reproduced from those figures, although we will not focus on comparing to data here.

We first compare the $50\hmpc$ and $100\hmpc$ \simba\ runs, which differ only in that the latter has $8\times$ the volume. In general they shows similar results within uncertainties, but the variance is larger in the $50\hmpc$ case owing to the fewer numbers of galaxies.  One notable difference is that the smaller volume produces a higher HIMF, which is within the variance at $z=0$ but not at $z=1$.  With a larger volume, there are more quenched galaxies with less \HI\ in their outskirts, resulting in a higher mass function for the smaller volume.  The difference is less noticeable for H$_2$, and in this case the large volume has a slightly higher mass function.  Nonetheless, these are all variations that are within $1\sigma$ expectations of cosmic variance, and since these two volumes are started from different initial conditions, such variations are not unexpected. 

We can examine resolution convergence by comparing the $50\hmpc$ (solid) and $25\hmpc$ (dotted) results, which have the same dynamic range but the latter has $8\times$ better resolution.  As such, it resolves galaxies to much lower masses.  For the H2MF, there is quite good resolution convergence, but it can be seen that the turn-down at $M_{\rm H_2}\la 10^9 {\rm M}_\odot$ owes primarily to incompleteness, as the $25\hmpc$ results do not show a turnover there.  The HIMF, in contrast, does tend to be slightly higher in the higher resolution volume. This general behavior is also seen between the main EAGLE run and the high-resolution EAGLE-Recal run~\citep{Crain:2017}, though with a more dramatic difference than in \simba.  However, in \simba, this might also be a volume effect as described in the previous paragraph.  More crucially, it is now evident that the turn-down in the HIMF in \simba\ clearly owes to numerical resolution -- with a high-resolution run, the HIMF continues to broadly follow the observed HIMF trend down to $\mhi\la 10^9{\rm M}_\odot$.  The impact of resolution is particularly evident at $z=1$, and suggests that the main \simba\ volume can only resolve \HI\ in galaxies down to $\mhi\sim 10^{10}{\rm M}_\odot$ at these redshifts. Upcoming surveys, however, are unlikely to probe to lower \HI\ masses, at least for individual detections.

Turning to the AGN feedback variants, at $z=0$ we see by far the strongest impact owes to the jet feedback mode.  This can be seen by noting that turning off X-rays (blue$\to$red) only modestly changes the cold gas mass functions.  Thus X-ray feedback does have some effect, but it is barely at the $1\sigma$ level relative to cosmic variance.  In contrast, further turning off jet feedback (red$\to$green) makes a much larger difference, which is true for both \HI\ and H$_2$.  \citet{Dave:2019} noted (via analogous tests) that it is jet feedback that is primarily responsible for quenching star formation in \simba, so it is unsurprising to see this reflected in H$_2$, but it is perhaps not immediately evident that it would also impact the \HI\ so dramatically.  Finally, we note that radiative AGN feedback has very little impact on cold gas content (green$\to$cyan).  If anything, including this feedback mode tends to increase the \HI, which may be because modest-velocity AGN winds are able to throw more material (albeit a small amount overall) into the CGM that can then remain sufficiently dense to be self-shielded.

Looking at $z=1$ (bottom panels), it is clear the differences are much less pronounced.  We do not show the results at $z=2$, but here all models are essentially identical to within cosmic variance. This indicates that the impact of jet feedback is mostly seen at $z\la 1$, which is also reflected in larger-scale motions of baryons in the IGM in \simba~\citep{Borrow:2019,Christiansen:2019}.  Without jets (green and cyan), the cold gas mass functions actually increase at the massive end from $z=1\to 0$, whereas with jets (blue and red) they decrease.  Hence in \simba, jet feedback that is responsible for quenching galaxies also dramatically changes late-time cosmic cold gas evolution.

Fig.~\ref{fig:gass_agn} gives a complementary view of the variations in cold gas properties induced by AGN feedback. This shows cold gas scaling relation comparisons at $z=0$ as in Fig.~\ref{fig:gass}, but here we compare the suite of $50\hmpc$ AGN feedback variant runs vs. xGASS and xCOLD~GASS scaling relations.  The blue, red, green, and cyan lines show running median scaling relations for the full \simba, No-Xray, No-jet, and No-AGN variants as described above.  Note that the full \simba\ run here is in a $50\hmpc$ box, rather than the fiducial $100\hmpc$ box shown in Figure~\ref{fig:gass}, in order to homogenise the comparison to the other runs.  The galaxies in the $50\hmpc$ volume have somewhat higher \HI\ contents and slightly lower H$_2$ contents than the $100\hmpc$ box, as seen in Figure~\ref{fig:mfgas}.  The observations from xGASS and xCOLD~GASS are reproduced from Fig.~\ref{fig:gass} for reference, though the focus in this plot is a comparison among the AGN feedback variants.

AGN feedback primary impacts high-mass galaxies in \simba. This is seen by the fact that at $M_*\la 10^{9.5}{\rm M}_\odot$, all models give similar results, even without any AGN feedback at all.  There is immediately an increase in gas content by turning on the radiative mode feedback (i.e. cyan No-AGN vs. green No-Jet lines), which as we discussed in the mass functions section seems counterintuitive:  Radiative AGN feedback increases the amount of both \HI\ and H$_2$ (as well as $L_{\rm CO}$).  This is because there is additional cold material being driven out of the galaxy that both mildly lowers the amount of stars formed while also providing more gas to accrete from the CGM.

The largest difference comes from further turning on jet feedback (green vs. red lines).  This causes a major drop in the gas fractions at $M_*\ga 10^{10}{\rm M}_\odot$, particularly in molecular gas.  It is this feedback, which is also responsible for quenching galaxies~\citep{Dave:2019} that is thus responsible for reproducing the strongly negative slope of gas fractions as observed; such a slope is not a trivial outcome of galaxy formation, but rather driven specifically by AGN feedback.  Note that the \HI\ fractions still have a negative slope even without jet feedback, since the growth of gravitational hot gaseous halos suppresses cold gas in the CGM relative to the stars; in contrast, the molecular fractions are nearly mass-independent without jet feedback.  Adding X-ray feedback has a small but noticeable impact on the molecular content of the most massive systems $M_*\ga 10^{11}{\rm M}_\odot$, which comes from the removal of molecular gas from the central regions owing to this feedback mode in \simba~\citep{Appleby:2019}.

The sSFR scaling relations (middle column) show much less dependence on AGN feedback.  This is primarily because the most significant impact of AGN feedback is to quench galaxies and build a population with very low sSFR; the ones that remain star-forming and dominate this plot are not strongly affected.  Nonetheless, once can see that turning on radiative mode AGN feedback, for instance, mildly increases the amount of \HI\ and H$_2$ in galaxies at modest sSFR.  The slopes of the sSFR--$f_{\rm gas}$ relations are thus in somewhat better agreement with observations when AGN feedback is included versus having all AGN feedback turned off.

The stellar surface density plot provides some insight into the nature of the discrepancies seen in Fig.~\ref{fig:gass} for the full \simba\ run.  Previously we argued that resolution may be impacting the results to prevent very high surface densities.  However, when jets are off (green/cyan lines), it is clear that plenty of galaxies at high stellar surface densities as observed can indeed be achieved at this numerical resolution.  Instead, it is primarily the jet feedback that causes the lack of high $\mu_*$ galaxies, with X-ray feedback providing a minor addition.  This is seen particularly dramatically in the $\fmol$ plot, where turning on X-ray feedback starkly reduces the number of high-$\mu_*$ galaxies. Note that \citet{Appleby:2019} found that this same X-ray feedback is necessary to obtain central depressions in the sSFR profiles of green valley galaxies in accord with observations, yet the impact on the stellar surface densities may be overly strong.  It could still be that the results are impacted by numerical resolution; simply producing high-$\mu_*$ galaxies does not guarantee that resolution is not a concern.  Nonetheless, this comparison illustrates that the stellar surface densities could potentially provide a complementary constraint on X-ray feedback, constrained by the internal buildup of stars within galaxies.

Overall, the comparison between AGN feedback variants in \simba\ demonstrates that the cold gas mass function and its evolution since $z\sim 1$ provides interesting constraints on AGN feedback models.  \simba's AGN feedback was designed to quench massive galaxies, while keeping the kinetic energy from the jets to be relatively modest, in accord with observations~\citep{Whittam:2018,Dave:2019}.  Given the large impact of this feedback mode in impacting the HIMF and H2MF, it may be likely that the remaining discrepancies versus e.g. the H2MF could be mitigated with modest tweaking of the jet energy input, while not substantively impact the quenched population.  \simba's jet AGN feedback has the largest impact on cold gas properties, and it may actually over-suppress high high stellar surface density systems.  Nonetheless, it is a non-trivial success that \simba, as well as other simulations, are able to come close to reproducing cold gas mass functions particularly at the massive end via AGN feedback, and highlights such data as a way to independently constrain a poorly understood aspect of modern galaxy formation models.

\subsection{Discussion: The interplay of feedback and cold gas}\label{sec:discussion}

We have seen that despite very similar stellar and SFR properties, EAGLE, \TNG, and \simba\ have substantially different predictions for cold gas properties.  In this section we briefly speculate on the physical origin of these model differences.  A proper study of this would require a systematic parameter space exploration of a range of feedback prescriptions, which is beyond the scope of this work, but there are some qualitative differences in feedback models that may broadly explain some of the variations.

At low masses, the dominant feedback mechanism in these simulations is star formation-driven outflows.  In both \simba\ and \TNG, star formation feedback is done via decoupled kinetic outflows, meaning they explicitly do not impact or remove ambient ISM gas.  In EAGLE, the feedback is done by raising the temperature of gas particles neighboring newly-formed star particles, by a temperature increment of $10^{7.5}$K. This drives outflows from the resulting pressure gradient, and also adds heat to the ISM, likely lowering the cold (both molecular and atomic) gas content. Also, it may heat surrounding CGM gas more, which would particularly lower the \HI\ content associated with the galaxy.  This may explain why, at low masses, EAGLE predicts lower \HI\ and particularly H$_2$ contents, indicating that the low-mass gas content of galaxies is sensitive to the amount of thermal energy deposition from star formation feedback into the ISM and CGM. This explanation seems plausible when one considers that the effect is strongly resolution-dependent, and that EAGLE's thermal feedback scheme injects less energy per feedback event at higher resolution.  

Larger differences are seen in the HIMF and H2MF at high masses above $\ga M^\star$.  In this regime, AGN feedback is an important contributor.  Again, \simba\ and \TNG\ have qualitatively similar AGN feedback approaches, via kinetic jets in massive galaxies at late times.  EAGLE continues to use a thermal-based feedback approach, with a higher temperature increment of $10^{8.5}$K for heating gas around the black hole.  Judging from the lack of massive cold gas reservoirs in EAGLE, it may be that this feedback mode is overly aggressive in heating or expelling cold gas. On the other hand, the overproduction of H$_2$ in \simba\ and \TNG\ (modulo uncertainties in $\alphaCO$) may indicate that a purely kinetic feedback form is insufficient to remove molecular gas from the ISM in a manner in accord with observations.  Note that \simba\ generally has weaker jet energy input than \TNG\ and does not vary the direction of the jet feedback on short timescales as \TNG\ does, but it does include X-ray feedback that is important for reducing the central molecular gas in quenching high-mass galaxies as observed~\citep{Appleby:2019}.  Thus the molecular gas content in massive galaxies appears to be impacted by essentially all the different feedback mechanisms, making detailed tests of the impact of each one within each simulation model (as in \S\ref{sec:simba_agn}) important for understanding their impact.  We leave such a study for future work.

Among the more striking results is that \TNG\ predicts comparable or lower \HI\ and H$_2$ mass functions at higher redshifts versus $z=0$, while \simba\ and EAGLE both have increasing mass function to high-$z$.  This is particularly curious since both \TNG\ and \simba\ used decoupled kinetic winds and two-mode AGN feedback, yet yield qualitatively different evolution.  One potential difference is that \TNG\ uses spherical thermal AGN feedback at high black hole Eddington fractions that are more common at earlier epochs, only going to kinetic at lower Eddington fractions, while \simba\ uses bipolar kinetic AGN feedback in all cases.  It could be that such thermal feedback during the peak of black hole growth activity is significantly impacting the gas in the immediate vicinity of star-forming galaxies, thereby modulating the cold gas content.  This does not happen in \simba\ owing to the explicit bipolar and decoupled nature of the AGN feedback, even at high Eddington fractions.  Since the largest impact is on the \HI, it is not immediately clear which approach is in better agreement with data since \HI\ mass functions aren't yet available at higher redshifts, but this is an interesting prediction that may already be testable using e.g. the upcoming LADUMA survey on MeerKAT.

\section{Summary}\label{sec:summary}

We have examined the atomic and molecular hydrogen properties of galaxies in three state-of-the-art cosmological hydrodynamic simulations, \simba, \TNG, and EAGLE, and compared them to available observations focusing on the $z\approx 0$ xGASS and xCOLD~GASS stellar mass-limited surveys.  \romeel{These simulations employ sub-grid models for self-shielding and H$_2$ formation.  In all models, the self-shielding is done via the prescription in \citet{Rahmati:2013}.  The H$_2$ fraction in \simba\ is computed on the fly using the \citet{Krumholz:2011} prescription, while EAGLE uses a similar method in post-processing, and TNG using a related prescription from \citet{Gnedin:2014}.}~
We have studied gas mass functions, CO(1-0) luminosity functions with $\alphaCO$ computed in the simulations using the prescription from \citet{narayanan12a}, and gas scaling relations versus stellar mass, specific star formation rate, and stellar mass surface density.  We also looked in \simba\ to examine how the cold gas properties vary when excluding individual AGN feedback processes.  We summarize our main results as follows.

\begin{itemize}
    \item \simba, EAGLE, and \TNG\ all have $z=0$ stellar mass functions and star-forming main sequences that are in good agreement with each other and with observations, indicating that stellar properties can now be well reproduced in the current generation of hydrodynamic models.
    \item The simulations' \HI\ mass functions all show a Schechter shape with a flat faint-end slope as observed, but the characteristic mass $M_{\rm HI}^\star$ varies substantially, from $2\times 10^9{\rm M}_\odot$ for EAGLE, to $10^{10}{\rm M}_\odot$ for \simba\ (in good agreement with ALFALFA data), to $2\times 10^{10}{\rm M}_\odot$ for \TNG.  EAGLE-Recal produces significantly more \HI\ than EAGLE, thus showing some resolution sensitivity for this model.
    \item \simba\ and EAGLE show a shift towards higher $M_{\rm HI}^\star$ at higher redshifts, increasing by $\sim\times 2-2.5$ out to $z=0$, while \TNG\ shows a dropping $M_{\rm HI}^\star$.  This qualitative difference will hopefully be discriminated in the upcoming generation of \HI\ surveys.
    \item The H$_2$ mass functions for these simulations likewise all show a flat faint-end slope as observed, but with dramatic variations in $M_{\rm H_2}^\star$, from $10^{9.2}{\rm M}_\odot$ for EAGLE to $10^{10}{\rm M}_\odot$ for \simba\ and \TNG; xCOLD~GASS finds $M_{\rm H_2}^\star\approx 10^{9.7}{\rm M}_\odot$, intermediate between these predictions.
    \item As with \HI, there is a strong increase in $M_{\rm H_2}^\star$ out to $z=2$ for EAGLE ($\sim\times 4$) and \simba\ ($\sim\times 2.5$), but no change for \TNG.  Thus the evolutionary differences are mainly in the total neutral gas content, not in the relative fractions of \HI\ and H$_2$.
    \item Examining the more directly-observable $L_{\rm CO(1-0)}$ rather than $M_{\rm H_2}$ gives a substantively different picture, because $\alphaCO$ can vary significantly for galaxies both within a model, and between models.  EAGLE and \simba\ produce on average $\alphaCO(z=0)\approx 3\aCOunits$ for $M_*>10^{10}{\rm M}_\odot$ star-forming galaxies, while \TNG\ has $\alphaCO(z=0)\approx 6\aCOunits$.  There is strong evolution in $\alphaCO$ to higher redshifts, with a median $\alphaCO(z=2)\approx 1-1.5\aCOunits$ in all models.
    \item The resulting simulated $z=0$ CO(1-0) luminosity functions are generally in closer agreement with xCOLD~GASS observations versus the H2MF, with EAGLE and \TNG\ agreeing very well and \simba\ still somewhat too high.
    \item The evolution of the COLF broadly mimics that of the H2MF.  EAGLE and \simba\ show much higher COLFs at $z=2$ than $z=0$.  These models well reproduce observations of high $L_{\rm CO1-0}$ galaxies at these epochs in the COLDz and ASPECS surveys, showing that at least some modern galaxy formation models can accommodate these observations.
    \item Assuming a constant $\alphaCO$ poorly approximates the $z=0$ COLF for EAGLE and \TNG, while for \simba\ it is a decent approximation at $z=0$ but not at $z=2$.  This shows that assuming a constant $\alphaCO$ can significantly bias H$_2$ comparisons, and that the exact way in which it will be biased depends in detail on the distribution of $\alphaCO$ within the simulated galaxy population. \romeel{Comparing model predictions using the \citet{Accurso:2017} $\alphaCO$ COLF vs. the \citet{narayanan12a} one, we find that the former produces a somewhat lower COLF at $z=0$, and a dramatically lower one at $z=2$ where its predicted $\alphaCO$ values are dubiously high.}
    \item Gas fractions drop with $M_*$ and rise with sSFR in all models, with slopes generally as seen in the data.  All models broadly reproduce the \HI\ and H$_2$ fractions but with some discrepancies such as \simba\ being too high at low masses, \TNG\ too high at high masses, and EAGLE slightly low overall, though EAGLE-Recal shows very good agreement.
    \item Comparing to $L_{\rm CO(1-0)}$ scalings, EAGLE produces reasonable agreement, \simba\ is too high particularly at low $M_*$ and sSFR, and \TNG\ may be too high at high $M_*$.
    \item Scaling relations versus stellar surface density $\mu_*$ generally show good agreement for \TNG\ and poorer agreement for EAGLE and \simba, which may in part reflect \TNG's superior numerical resolution for modeling the stellar surface density, but may also reflect failings of the feedback models.  EAGLE-Recal shows higher $\mu_*$ values than EAGLE, which  corroborates at least some of the discrepancy owing to numerical resolution.
    \item Comparing AGN feedback variants in \simba\ shows that it is \simba's jet mode feedback that is responsible for suppressing cold gas mass functions at the massive end from $z\sim 1\to 0$; without jet feedback, the mass functions grow with time.  
    \item Jet AGN feedback in \simba\ is further responsible for creating a strong anti-correlation between molecular gas fraction and stellar mass; without this, \simba\ predicts roughly constant molecular gas fractions.  Concurrently, jet feedback seems to over-suppress the formation of the highest stellar surface density objects, which could be responsible for \simba's discrepancies versus these observations.
    \item \romeel{Besides $\alphaCO$ and numerical resolution, the dominant systematic for comparisons to observations is the choice of aperture, particularly for H$_2$ in massive galaxies. While this might be partly mitigated by carefully mocking observations, the H$_2$ formation threshold and ISM pressurization used in all these simulations intrinsically limit how well the ISM structure can be modeled at cosmological resolutions.  Aperture effects are unlikely to qualitatively change our results, but for robust quantitative predictions it will be necessary to model such systematics more carefully.}
\end{itemize}

These results illustrate how cold gas content and its evolution provides strong constraints on key galaxy growth and feedback processes in current models.  Even just the mass functions of atomic and molecular gas and their evolution are already qualitatively different in our state-of-the-art simulations that have very similar stellar properties.  While \HI\ provides a more straightforward comparison, current radio telescopes provide no constraints at $z\gg 0$, although this will hopefully improve soon with now-online Square Kilometre Array (SKA) precursors MeerKAT and ASKAP.  The molecular gas content is potentially even more constraining, but current comparisons are strongly systematics-limited in terms of assumptions regarding $\alphaCO$ that connects the simulated H$_2$ mass to CO luminosity. More attention should be given to modeling this quantity, which may depend sensitively on many sub-grid aspects of ISM modeling.  Nonetheless, the large differences between the evolutionary trends in models suggests that even broad constraints on $\alphaCO$ could provide substantial discriminatory power.  This would unlock the full potential of ALMA data to employ the evolution of molecular gas as a constraint on galaxy formation models.  The next generation of far-infrared and radio facilities promises to provide novel and complementary constraints on galaxy formation models that will be crucial for building a more comprehensive scenario of galaxy evolution within a cosmological context.

\section*{Acknowledgements}

The authors acknowledge helpful discussions with Daniel Angl\'es-Alc\'azar, Sarah Appleby, Katarina Kraljic, and Daniele Sorini, as well as helpful comments from the referee.  We thank Robert Thompson for developing {\sc Caesar}, and the {\sc yt} team for development and support of {\sc yt}.
RD acknowledges support from the Wolfson Research Merit Award program of the U.K. Royal Society. RAC is a Royal Society University Research Fellow. ARHS is the Jim Buckee Fellow at UWA. DN is supported in part by NSF Award AST-1715206 and HST Theory Award 15043.0001.  LC is the recipient of an Australian Research Council Future Fellowship (FT180100066) funded by the Australian Government.
This work used the DiRAC@Durham facility managed by the Institute for Computational Cosmology on behalf of the STFC DiRAC HPC Facility. The equipment was funded by BEIS capital funding via STFC capital grants ST/P002293/1, ST/R002371/1 and ST/S002502/1, Durham University and STFC operations grant ST/R000832/1. DiRAC is part of the National e-Infrastructure.
Parts of this research were conducted by the Australian Research Council Centre of Excellence for All Sky Astrophysics in 3 Dimensions (ASTRO 3D), through project number CE170100013.  \romeel{This research was supported by the Munich Institute for Astro- and Particle Physics (MIAPP) which is funded by the Deutsche Forschungsgemeinschaft (DFG, German Research Foundation) under Germany's Excellence Strategy -- EXC-2094-390783311.}

\section*{Data Availability Statement}

The data underlying this work, including simulation snapshots and galaxy catalogs, will be shared on reasonable request to the corresponding author R. Dav\'e.

\bibliographystyle{mnras}
\bibliography{coldgas} 


\bsp	
\label{lastpage}
\end{document}